\documentclass[aip,jcp,
reprint,%
floatfix]{revtex4-1}

\usepackage[usenames,dvipsnames]{color}
\usepackage[english]{babel}
\usepackage{epsfig}
\usepackage{float}
\usepackage{graphicx,subfigure}
\usepackage{tikz}
\usepackage{pgfplots}
\pgfplotsset{compat=1.12}
\usetikzlibrary{calc}
\usetikzlibrary{decorations.markings}
\usetikzlibrary{decorations.pathmorphing}
\usetikzlibrary{positioning,shapes,arrows,decorations.markings,calc}
\usepackage{multirow}
\usepackage{arydshln}

\usepackage{amsmath,amssymb}
\usepackage[mathcal,mathscr]{eucal}
\usepackage[utf8]{inputenc}
\usepackage[T1]{fontenc}
\usepackage[english]{babel}
\usepackage{times}
\usepackage{lmodern}
\usepackage[Euler]{upgreek}
\usepackage{enumerate}
\usepackage[colorlinks=true, allcolors=blue]{hyperref}
\usepackage{mathptmx}
\usepackage{bookmark}

\usepackage{textcomp}
\usepackage{stmaryrd}

\usepackage{xr}
\makeatletter
\newcommand*{\addFileDependency}[1]{
  \typeout{(#1)}
  \@addtofilelist{#1}
  \IfFileExists{#1}{}{\typeout{No file #1.}}
}
\makeatother

\newcommand*{\myexternaldocument}[1]{
    \externaldocument{#1}
    \addFileDependency{#1.tex}
    \addFileDependency{#1.aux}
}
\myexternaldocument{SM}

\setcitestyle{numbers,square,citesep={,\kern-.24em}}
\addto\captionsenglish{\renewcommand{\figurename}{FIG.}}
\addto\captionsenglish{\renewcommand{\tablename}{TABLE}}

\begin{document}

\preprint{AIP/123-QED}

\title
{Atomic and electronic structure of cesium lead triiodide surfaces}

\author{Azimatu Seidu}
\affiliation{Department of Applied Physics, Aalto University, FI-00076 AALTO, Finland}
\author{Marc Dvorak}
\affiliation{Department of Applied Physics, Aalto University, FI-00076 AALTO, Finland}
\author{Patrick Rinke}
\affiliation{Department of Applied Physics, Aalto University, FI-00076 AALTO, Finland}
\author{Jingrui Li}
\affiliation{Electronic Materials Research Laboratory, Key Laboratory of the Ministry of Education \& International Center for Dielectric Research, School of Electronic Science and Engineering, Xi'an Jiaotong University, Xi'an 710049, China}

\begin{abstract}
The (001) surface of the emerging photovoltaic material cesium lead triiodide (CsPbI$_3$) is studied. Using first-principles methods, we investigate the atomic and electronic structure of cubic ($\upalpha$) and orthorhombic ($\upgamma$) CsPbI$_3$. For both phases, we find that CsI-termination is more stable than PbI$_2$-termination. For the CsI-terminated surface, we then compute and analyse the surface phase diagram. We observe that surfaces with added or removed units of nonpolar CsI and PbI$_2$ are most stable. The corresponding band structures reveal that the $\upalpha$ phase exhibits surface states that derive from the conduction band. The surface reconstructions do not introduce new states in the band gap of CsPbI$_3$, but for the $\upalpha$ phase we find additional surface states at the conduction band edge. 
\end{abstract}

\maketitle

\section{Introduction}\label{intro}

In recent years, perovskite solar cells (PSCs) have generated increased attention within the photovoltaic community. The most common PSC photoabsorbers are hybrid organic-inorganic halide perovskites (HPs) with ABX$_3$ structure, where  A is an (organic) monovalent cation, B either Pb or Sn, and X a halogen. Among the HPs, the most widely studied materials are methylammonium (MA) lead iodide ($\text{CH}_3^{}\text{NH}_3^{}\text{PbI}_3^{}$ or $\text{MAPbI}_3^{}$) and formamidinium (FA) lead iodide [$\text{HC}(\text{NH}_2^{})_2^{}\text{PbI}_3^{}$ or $\text{FAPbI}_3^{}$]. HPs are the most promising materials for next-generation photovoltaic technologies as reflected by their rapidly rising power conversion efficiency (PCE): it reached $\sim$ 25\% \cite{NRELchart19} only seven years after the invention of the state-of-the-art PSC architecture in 2012 (PCE $\sim$10\%) \cite{kim, lee}. HPs are also promising for light emitting diodes, lasers, and photodetectors \cite{Huang14, Lin15, Cho15}. Their outstanding properties for optoelectronic applications include optimal band gaps, excellent absorption in the visible range of the solar spectrum, exceptional transport properties for both electrons and holes, flexibility of composition engineering, as well as low cost in both materials and fabrication \cite{snaith, green, stranks, Xing13, Eperon15, troughton}.

Despite the rapid PCE improvement in the laboratory, stability issues limit the development and commercialization of HPs for real photovoltaic applications. Especially, the organic components in hybrid perovskites are susceptible to ambient conditions such as moisture, oxygen, and heat, and exposure leads to rapid performance degradation \cite{NiuG14,NiuG15,HuangJ17,KimGH17,Mesquita18,Ciccioli18,LiF18}. 
Several approaches have been proposed to solve these pressing stability problems, including surface protection with organic long-chain ligands \cite{SchmidtL14, SoranyelG15,Dong19}, synthesis of quasi two-dimensional perovskites \cite{Quan16,Dou17,Ran18a,WangZ18,LiuC18,Ran19}, protective coating with inorganic semiconductors or insulators \cite{Matteocci16,Cheacharoen18a,Cheacharoen18b, Seidu19}, and A-site  substitution with smaller monovalent ions \cite{noh,Yi16,ZhouY16,Tan17,Ciccioli18,Gao18, Baena17, Ganose17}. 

In the context of A-site substitutions, the all-inorganic perovskite $\text{CsPbI}_3^{}$ and its mixed-halide derivatives have emerged as a promising alternative to the hybrid MA- and FA-based perovskites. $\text{CsPbI}_3^{}$ has a similar structure and slightly closer Pb--I packing and higher thermal and chemical stabilities than MAPbI$_3$ and FAPbI$_3$ \cite{Eperon15, Forolova17}. The latest PCE of $\text{CsPbI}_3^{}$-based PSCs has already reached 18\% \cite{WangY19}, but more materials design and device engineering are needed to increase the conversion efficiency and the operational stability. This applies to several aspects, such as morphology control of the HP thin films, interface engineering between $\text{CsPbI}_3^{}$ and interlayer materials, and the passivation of intrinsic defects at the interfaces and grain boundaries, which act as nonradiative recombination centers thus degrading the device efficiency. A comprehensive understanding of the atomic and electronic structure of $\text{CsPbI}_3^{}$ surfaces would advance its development as a PSC photoabsorber. The surfaces of MA- and FA-based perovskites have been investigated theoretically \cite{Haruyama14,Haruyama16, Oscar16, Akbari17} and experimentally \cite{Jiang19, Chen18, Cho18, Saliba16a, Saliba16b, Saliba16c}.
For $\text{CsPbI}_3^{}$\,, however, we are only aware of bulk defect studies \cite{Li17, Huang18, Sutton18, Evarestov19}. The surfaces and interfaces of $\text{CsPbI}_3^{}$ have not yet been considered.

In this work, we present first-principles density functional theory (DFT) calculations for the reconstructed surfaces of the photovoltaic-active $\upalpha$ (cubic) and $\upgamma$ (orthorhombic) phases of $\text{CsPbI}_3^{}$\,. Starting from the pristine (clean) surface models with CsI- and PbI$_2$-terminations (denoted by CsI-T and PbI$_2$-T, respectively), constituent elements (Cs, Pb, and I) as well as their complexes (CsI, PbI, and PbI$_2$) were added to or removed from the surface. The thermodynamic stability of these surface models was investigated for different chemical environments by means of \textit{ab initio} thermodynamics \cite{karsten, reuter, abinitiothermodynamics}. For the stable surface models, we calculated their electronic structure and elucidated changes in their electronic properties in comparison to the clean surfaces. 

The remainder of this paper is organized as follows. In Sec.~\ref{methods}, we briefly outline the computational details of our DFT calculations  and summarize the thermodynamic constraints for the growth of bulk CsPbI$_3$ as well as the CsI-T and PbI$_2$-T surfaces. In Sec.~\ref{results}, we first analyze the stability of the clean-surface models (CsI-T and Pb$_2$-T) and the reconstructed models with missing- and add-atoms (and their complexes). We then discuss changes in crystal structures due to missing- and add-atoms with focus on their stability and their atomic and electronic structure. Finally, we conclude with a summary in Sec.~\ref{conclusion}.

\section{Computational details}\label{methods}

All DFT calculations were performed using the Perdew-Burke-Ernzerhof exchange-correlation functional for solids (PBEsol) \cite{perdew08} implemented in the all-electron numeric-atom-centered orbital code \textsc{fhi-aims} \cite{Blum09,havu,Levchenko/etal:2015}. We chose PBEsol because it describes the lattice constants of $\text{CsPbI}_3^{}$ well at moderate computational cost \cite{Yang17, Bokdam17}. Scalar relativistic effects were included by means of the zeroth-order regular approximation \cite{vanlenthe}. We used standard \textsc{fhi-aims} tier~2 basis sets for all calculations,  
in combination with a $\Gamma$-centered $4\times4\times4$ and a $4\times4\times1$ $k$-point mesh for the bulk materials and the surface calculations with a slab model, respectively. The bulk structures were optimized with the analytical stress tensor \cite{knuth}. 
For the slab models, we fixed the the lattice constants and all atomic positions except for atoms in the top and bottom CsPbI$_3$ units (the surface atoms). Surface-dipole correction \cite{Neugebauer92} was used in all surface calculations.

\subsection{Structural optimization}

\subsubsection{Bulk and surface structures}

\begin{figure}[!htp]
\includegraphics[width=\linewidth]{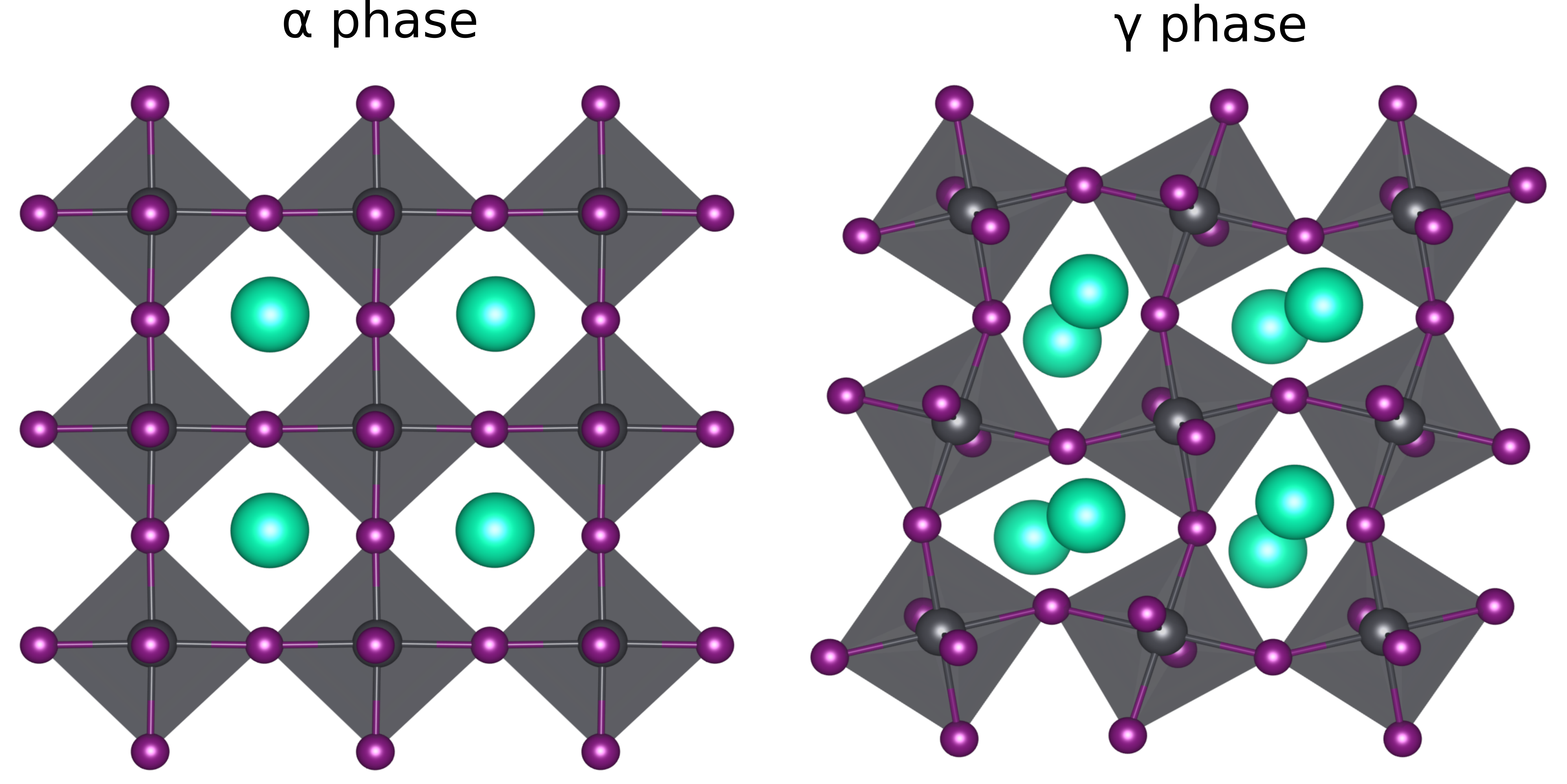}\vspace{-1.0em}
\caption{Bulk crystal structures of the $\upalpha$ (cubic) and the $\upgamma$ (orthorhombic) phases of $\text{CsPbI}_3^{}$\,. $\text{Cs}$, $\text{Pb}$, and $\text{I}$ are colored in green, black, and purple, respectively. The PbI$_6$ octahedra are colored in dark gray.
}\label{bulk}
\end{figure}

In this work, we considered two experimentally accessible photovoltaic-active perovskite phases of $\text{CsPbI}_3^{}$: the $\upalpha$ (cubic) phase with space group Pm\=3m and the $\upgamma$ (orthorhombic) phase with space group Pnma. For each phase, we constructed and optimized a $2\times2\times2$ bulk supercell with DFT (structures shown in Fig.~\ref{bulk}). The lattice parameters of our optimized $\upalpha$ phase are $a=b=c=12.47$~{\AA} and $\alpha=\beta=\gamma = 90\text{\textdegree}$. For the $\upgamma$ phase, the lattice parameters  are $a=b = 12.21$~{\AA}, $c = 12.35$~{\AA} and $\alpha = \beta = 90\text{\textdegree}$, $\gamma = 85.8\text{\textdegree}$.

\begin{figure}[!htp]
\includegraphics[width=\linewidth]{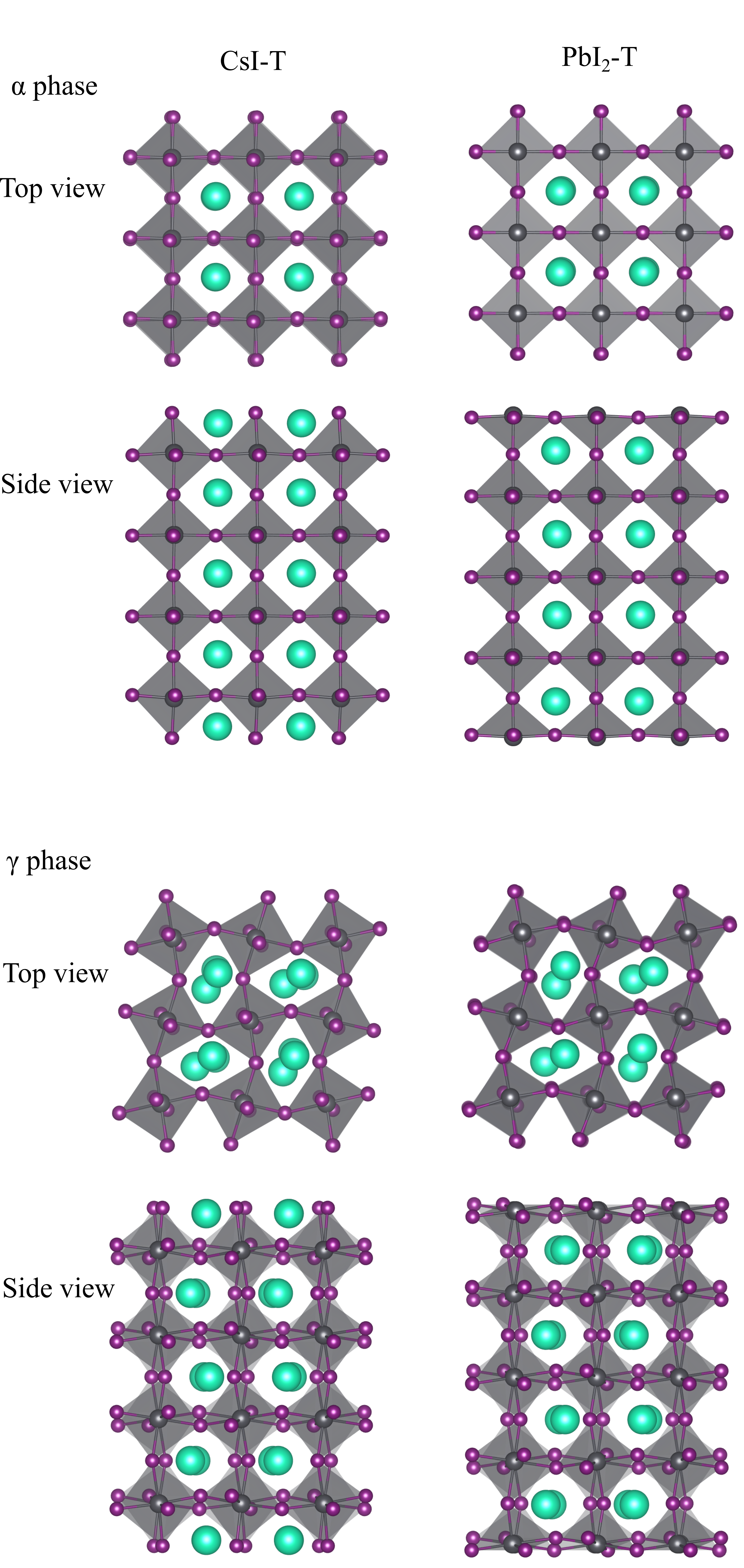}\vspace{-1.0em}\caption{Relaxed $\text{CsI}$-T and $\text{PbI}_2^{}$-T clean-surface models of the $\upalpha$ and the $\upgamma$ phases. Depicted on the left is the CsI-T and on the right the $\text{PbI}_2^{}$-T termination.}\label{surfaces}
\end{figure}

For each phase, we constructed the surface models by inserting a vacuum region in the $[001]$ direction of the investigated system. With a $30~\text{\AA}$ vacuum thickness and the inclusion of surface-dipole correction \cite{Neugebauer92}, we minimized the interaction between neighboring slabs.  
In this work, we focused on the $(001)$ surfaces which are the major facet of halide perovskites \cite{Haruyama14,Haruyama16,Schulz19} and the most relevant surfaces of $\text{CsPbI}_3^{}$\,. We carried out DFT calculations for $\text{CsPbI}_3^{}$ surfaces with symmetric slab models for CsI-T and PbI$_2$-T surfaces. As depicted in Fig.~\ref{surfaces}, the CsI-T surface model consists of 5 CsI and 4 PbI$_2$ layers alternatively stacked along the $[001]$ direction. Similarly, the PbI$_2$-T surface model has 5 PbI$_2$ and 4 CsI alternating layers.To avoid quantum confinement in our slab model band structure plots, we added 5 units of CsPbI$_3$ along the (001) direction of our 2$\times2$ supercell and calculated the band structures before and after relaxation.

For both CsI-T and $\text{PbI}_2^{}$-T surface models, we studied different missing- and add-atom reconstructions. The missing- and add-atoms are labeled as v$_{\text{X}}$ and i$_{\text{X}}$\,, respectively, with X indicating the atoms or their complexes. All add-atoms and their complexes were added to the surfaces, while missing-atoms were removed from the topmost layers containing those atoms. For instance, v$_{\text{Cs}}$\,, v$_{\text{I}}$\,, and v$_{\text{CsI}}$ of $\text{CsI}$-T surfaces were constructed by removing atoms from the topmost CsI layer while v$_{\text{Pb}}$\,, v$_{\text{PbI}}$\,, and v$_{\text{PbI}_2^{}}$ indicate the removal of atoms from the PbI$_2$ layer below the topmost CsI layer.  

The $2\times2$ surface unit cell allows us to study 26 reconstructed surface models each for CsI-T and PbI$_2$-T. In detail, these amount to 13 missing-atom or missing-complex models (v$_{\text{Cs}}$\,, v$_{2\text{Cs}}$\,, v$_{4\text{Cs}}$ [only in CsI-T], v$_{\text{Pb}}$\,, v$_{2\text{Pb}}$\,, v$_{4\text{Pb}}$ [only in PbI$_2$-T], v$_{\text{I}}$\,,  v$_{2\text{I}}$\,, v$_{\text{CsI}}$\,, v$_{2\text{CsI}}$\,, v$_{\text{PbI}}$\,, v$_{2\text{PbI}}$\,, v$_{\text{PbI}_2^{}}$\,, and v$_{2\text{PbI}_2^{}}$) and 13 add-atom or add-complex structures (i$_\text{Cs}$\,, i$_{2\text{Cs}}$\,, i$_{\text{Pb}}$\,, i$_{2\text{Pb}}$\,, i$_{\text{I}}$\,, i$_{2\text{I}}$\,, i$_{\text{CsI}}$\,, i$_{2\text{CsI}}$\,, i$_{4\text{CsI}}$ [only in CsI-T], i$_{\text{PbI}}$\,, i$_{2\text{PbI}}$\,, i$_{\text{PbI}_2^{}}$\,, i$_{2\text{PbI}_2^{}}$\,, and i$_{4\text{PbI}_2^{}}$ [only in  PbI$_2$-T]).  For double missing- and add-atoms (i.e., v$_{2\text{X}}^{}$ and i$_{2\text{X}}^{}$), we considered both line and diagonal options (i.e., removing $2$ iodine atoms along the $[100]$ or $[110]$ directions for v$_{2\text{I}}^{}$). We found no significant total-energy differences between these two modes. Hence we only present results from the diagonal modes in this paper. 

In pursuit of open materials science \cite{Himanen2019}, we made the results of all relevant calculations available on the Novel Materials Discovery (NOMAD) repository \cite{NOMAD}.

\subsection{Grand potential analysis}

For a system in contact with a particle reservoir and neglecting finite temperature contributions, the thermodynamic stability of a structure is obtained from the grand potential, $\varOmega$,
\begin{equation}
\varOmega \approx E - \sum_i x_i \mu_i \, . 
\end{equation}
Here $\mu_i$ is the chemical potential of species $i$ and $x_i$ the number of atoms of this species in the structure. The sum over $i$ runs over all elements in the compound. The relative stability between two systems both in contact with the same particle reservoir is determined by differences in $\varOmega$, with $\varOmega^A < \varOmega^B$ indicating that phase $A$ is more stable than phase $B$.
A special case of $\varOmega$ is when a system is in contact with its constituent species in their most stable phase. This defines the standard formation energy which is denoted by $\Delta H$ hereafter:
\begin{equation}
\Delta H = E - \sum_i x_i \mu_i^{\minuso} \, .
\end{equation}
Here, $\mu_i^{\minuso}$ indicates the chemical potential of species $i$ in its most stable form. The thermodynamic stability condition $\Delta H < 0$ states that the system's total energy must be lower than the sum of its constituents' chemical potentials, each in their most stable phase.

The chemical potentials $\mu_i$ are set by environmental conditions. We apply a simple transformation to the chemical potentials,
\begin{equation}
\Delta \mu_i = \mu_i - \mu_i^{\minuso} \, ,
\end{equation}
to introduce the parameter $\Delta \mu_i$\,. $\Delta \mu_i$ is the change in the chemical potential away from its value in the element's most stable phase, $\mu_i^{\minuso}$\,. The $\Delta \mu_i$ represent environmental growth conditions and are a convenient parameter to vary in order to map phase diagrams. The grand potential can be rewritten as
\begin{equation}
\varOmega = E - \sum_i x_i \mu_i^{\minuso} - \sum_i x_i \Delta \mu_i \, .
\end{equation}

The relative stability condition between phases A and B is then
\begin{eqnarray}
\varOmega^{A} &<& \varOmega^{B} 
\end{eqnarray}
which can be rearranged as
\begin{eqnarray}
\bigg( E^{A} -\sum_i x_i^{A} \mu_i^{\minuso} \bigg) &-& \bigg( E^{B} -\sum_i x_i^{B} \mu_i^{\minuso} \bigg)  \nonumber  \\
< \sum_i ( x_i^{A} &-& x_i^{B} ) \Delta \mu_i \, . \nonumber 
\end{eqnarray}
We recognize $\Delta H $ for phases $A$ and $B$,
\begin{equation}
\Delta H^{A} - \Delta H^B <  \sum_i ( x_i^{A} - x_i^{B} ) \Delta \mu_i \, . \label{inequality}
\end{equation}
The inequality given in Eq.~(\ref{inequality}) is the basis for the phase diagram, including the SPDs in this work. We calculate $\Delta H$ using the DFT total energy of the surface for $E$ and the DFT total energy per species unit for $\mu_i^{\minuso}$\,. For the specific case of a surface formation energy, $\Delta H$ reduces to
\begin{equation}
\Delta H^{\mathrm{surf}} = E^{\mathrm{surf}} - E^{\mathrm{bulk}} - E^{\mathrm{ads}} \label{surf_h}
\end{equation}
for surface total energy $E^{\mathrm{surf}}$, bulk total energy $E^{\mathrm{bulk}}$, and the total energy of any adsorbants $E^{\mathrm{ads}}$. With the various total energies tabulated from DFT calculations, we plot an SPD based on the inequalities in Eq.~(\ref{inequality}) as a function of the parameters $\Delta \mu_i$\,.

\subsection{Thermodynamic constraints for stable \texorpdfstring{$\text{CsPbI}_3^{}${}}~~bulk and surfaces}\label{therm_stab}

We first consider conditions for stable $\text{CsPbI}_3^{}$ in the bulk. In order to avoid the formation of atomic Cs, Pb, and I as well as bulk CsI and $\text{PbI}_2^{}$\,, the region of the phase diagram for stable $\text{CsPbI}_3^{}$ is determined by the inequalities,
\begin{equation}
\begin{split}
\Delta H (\text{CsPbI}_3^{}) &\leqslant \Delta\mu_{\text{Cs}}\leqslant 0 \, , \\
\Delta H (\text{CsPbI}_3^{}) &\leqslant \Delta\mu_{\text{Pb}}\leqslant 0 \, , \\
\Delta H (\text{CsPbI}_3^{}) &\leqslant 3 \Delta\mu_{\text{I}}\leqslant 0 \, ;
\end{split} \label{stab4}
\end{equation}
and
\begin{equation}
\begin{split}
\Delta H (\text{CsPbI}_3^{}) &\leqslant \Delta\mu_{\text{Cs}} + \Delta\mu_{\text{Pb}} + 3\Delta\mu_{\text{I}} \, , \\
\Delta\mu_{\text{Cs}} +  \Delta\mu_{\text{I}} &\leqslant \Delta H (\text{CsI}) \, , \\
\Delta\mu_{\text{Pb}} + 2\Delta\mu_{\text{I}} &\leqslant \Delta H (\text{PbI}_2^{}) \, .
\end{split} \label{stab2}
\end{equation}
The inequalities in Eq.~(\ref{stab2}) can be rearranged as
\begin{equation}
\begin{split}
\Delta H (\text{CsPbI}_3^{}) &\leqslant \Delta\mu_{\text{Cs}} + \Delta\mu_{\text{Pb}} + 3\Delta\mu_{\text{I}}  \, , \\
\Delta H (\text{CsPbI}_3^{}) - \Delta H (\text{CsI})
&\leqslant \Delta\mu_{\text{Pb}} + 2 \Delta\mu_{\text{I}} \leqslant \Delta H (\text{PbI}_2^{}) \, , \\
\Delta H (\text{CsPbI}_3^{}) - \Delta H(\text{PbI}_2)
&\leqslant \Delta\mu_{\text{Cs}} + \Delta\mu_{\text{I}} \leqslant \Delta H (\text{CsI}) \, .
\end{split} \label{stab3}
\end{equation}
$\mu_{\text{Cs}}^{\minuso}$\,, $\mu_{\text{Pb}}^{\minuso}$\,, $\mu_{\text{I}}^{\minuso}$ are calculated for the stable structures of Cs (I\=43m), Pb (P6$_3$/mmc), and I ($\text{I}_2^{}$ molecule). Eqs.~(\ref{stab4}) and (\ref{stab3}) are the conditions for stable $\text{CsPbI}_3$\,. Formation energies $\Delta H$ for Eqs.~(\ref{stab4}) and (\ref{stab3}) are calculated with DFT. Varying the three parameters $\Delta \mu_{\mathrm{Cs}}$\,, $\Delta \mu_{\mathrm{Pb}}$\,, and $\Delta \mu_{\mathrm{I}}$ maps the bulk stability region.

To compare the stability of two surfaces, we solve Eq.~(\ref{inequality}) to obtain the SPDs. Eq.~(\ref{inequality}) is a condition for surface stability in addition to Eqs.~(\ref{stab4}) and (\ref{stab3}), which are only for the bulk. The bulk and surface are not in isolation from each other. For this reason, the final surface stability is determined by overlaying the SPD on the bulk stability region. We consider the overlap of the stable bulk region with the SPD to be the predictor of a viable bulk and surface together.

\section{Results and Discussion}\label{results}

In this section, we present the results from our thermodynamic analysis, compare the stability of our surface termination models (CsI-T vs. PbI$_2$-T), and analyse the most relevant terminations using SPDs. We conclude the section with the electronic properties of the bulk and most relevant reconstructed surface models.

\subsection{Thermodynamic stability limits for \texorpdfstring{$\text{CsPbI}_3^{}${}}~~bulk and surface terminations}\label{thermo}

The PBEsol-calculated formation energies of bulk CsI, PbI$_2^{}$\,, $\upalpha$-CsPbI$_3^{}$\,, and $\upgamma$-CsPbI$_3^{}$ are $-3.40$, $-2.47$, $-5.89$, and $-6.02$~eV, respectively. From Eq.~(\ref{stab3}), the thermodynamic growth limits for bulk $\text{CsPbI}_3^{}$ in the $\upalpha$ and the $\upgamma$ phases at $\Delta\mu_{\text{Cs}} = 0 $ then are
\begin{equation}
\begin{split}
-2.49~\text{eV} \leqslant \Delta\mu_{\text{Pb}} + 2\Delta\mu_{\text{I}} &\leqslant -2.47~\text{eV} \quad \text{for } \upalpha, \\
-2.62~\text{eV} \leqslant \Delta\mu_{\text{Pb}} + 2\Delta\mu_{\text{I}} &\leqslant -2.47~\text{eV} \quad \text{for } \upgamma.
\end{split} \label{stab6}
\end{equation}
Similarly, the growth limits for bulk $\text{CsPbI}_3^{}$ at $\Delta\mu_{\text{Pb}} = 0 $ in the $\upalpha$ and the $\upgamma$ phases are
\begin{equation}
\begin{split}
-3.42~\text{eV} \leqslant \Delta\mu_{\text{Cs}} + \Delta\mu_{\text{I}} &\leqslant -3.40~\text{eV} \quad \text{for } \upalpha, \\
-3.55~\text{eV} \leqslant \Delta\mu_{\text{Cs}} + \Delta\mu_{\text{I}} &\leqslant -3.40~\text{eV} \quad \text{for } \upgamma.
\end{split} \label{stab7}
\end{equation}
The small difference between the left and the right values of these inequalities indicates the narrow stability region of bulk $\mathrm{CsPbI}_3^{}$\,. The stability window in the $\upalpha$ phase is especially small, only $\sim 0.02$~eV. For each phase, the width of this region equals the energy required for $\text{CsPbI}_3^{}$ to decompose into CsI and PbI$_2^{}$\,. Therefore, the narrow energy range for the growth of bulk $\text{CsPbI}_3^{}$ reflects the instability and ease of dissociation of $\mathrm{CsPbI}_3^{}$ into CsI and PbI$_2$\,, as alluded to in Sec.~\ref{intro}. 

\begin{figure}[!htp]
\includegraphics[width=\linewidth]{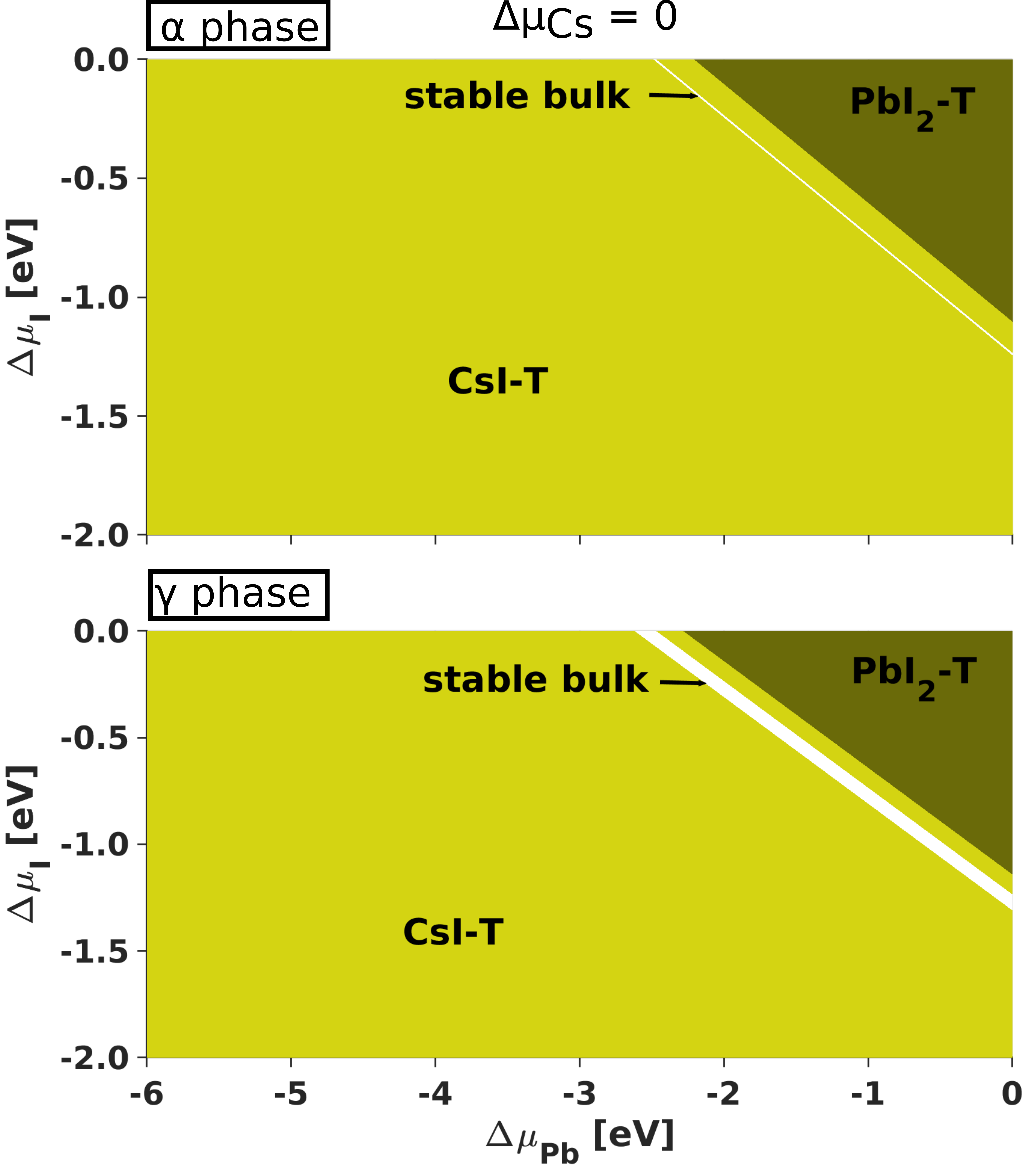}\vspace{-1.0em}
\caption{Thermodynamic growth limit for CsI-T and PbI$_2$-T surfaces in the $\upalpha$ and the $\upgamma$ phases at $\Delta\mu_{\mathrm{Cs}} = 0$. The white shaded regions depict the thermodynamically stable range for the growth of bulk $\text{CsPbI}_3^{}$\,. 
} 
\label{stab_reg_surf}
\end{figure}

Figure~\ref{stab_reg_surf} depicts the SPDs for the CsI-T and PbI$_2$-T clean surfaces in the $\upalpha$ and the $\upgamma$ phases at $\Delta\mu_{\mathrm{Cs}} = 0$. The stable bulk region is represented with the white shading. The CsI-T and PbI$_2$-T surfaces are stable in different regions. Since the CsI-T surface intersects the stable bulk region, we consider it more stable in conditions for bulk growth. Additionally, we observe stable CsI-T surfaces across a wider range of $\Delta\mu_k$ ($k =$ Cs, Pb, I) than PbI$_2$-T surfaces. The results of Fig.~\ref{stab_reg_surf} are similar to the findings of previous theoretical studies for $\text{MAPbI}_3^{}$ \cite{Yin14, Haruyama14,Haruyama16, Geng15} on the stability of methylammonium-iodide terminated over PbI$_2$-T surfaces. Our discussions will therefore focus on CsI-T surfaces from here on. Data for PbI$_2$-T surfaces including the relaxed surface-reconstruction structures and the SPDs are given in Supplementary Material (SM).

\subsubsection{Surface phase diagrams of CsI-T surface models}\label{surface_models}

\begin{figure*}[!htp]
\centering
\includegraphics[width=\textwidth]{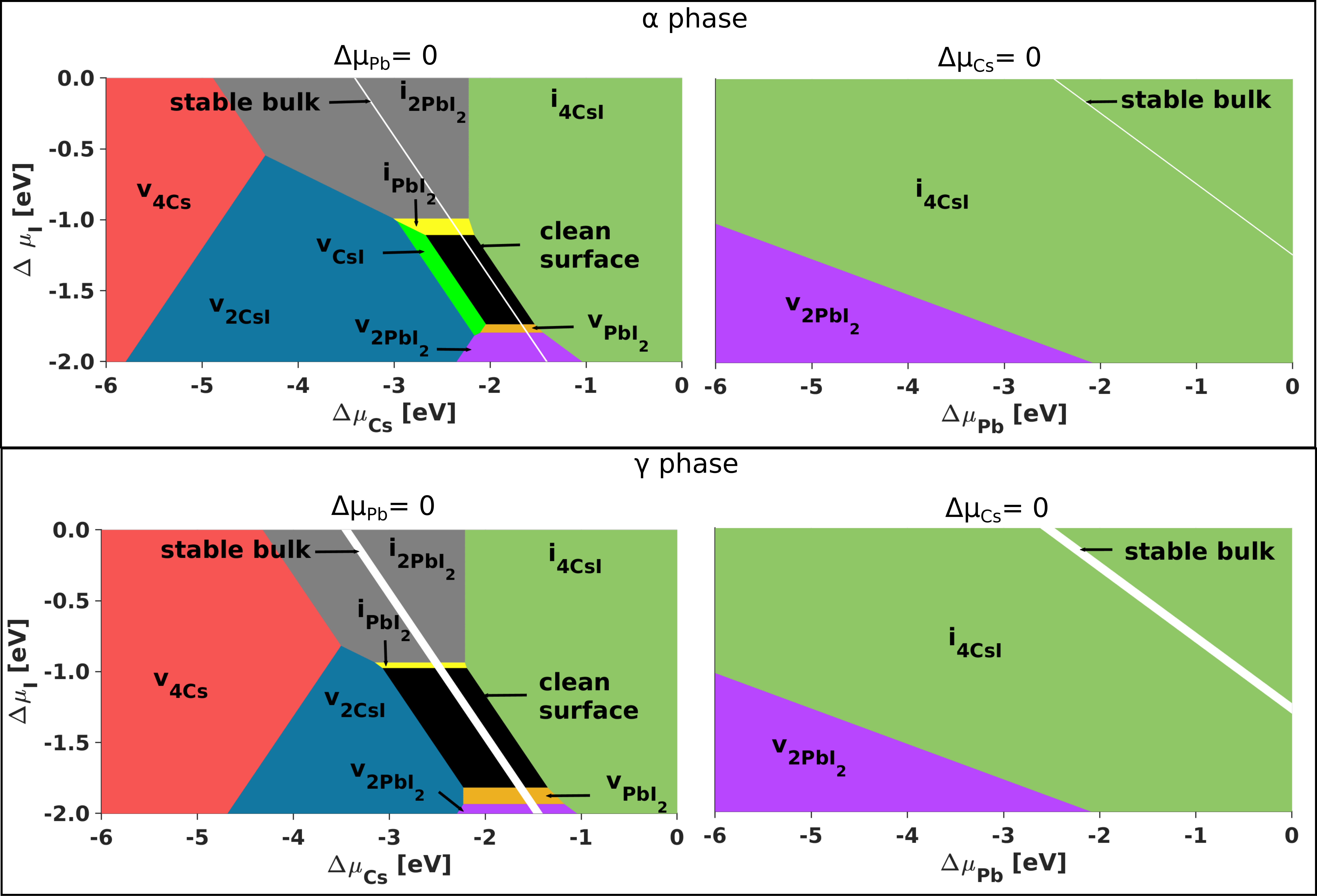}\vspace{-1.0em}
\caption{Surface phase diagrams of reconstructed CsI-T surfaces with missing- and add-atoms as well as their complexes (upper panel for the $\upalpha$ phase and lower panel for the $\upgamma$ phase). The white regions depict the thermodynamically stable range for the growth of bulk $\text{CsPbI}_3^{}$\,.}\label{spds_1}
\end{figure*}

Figure~\ref{spds_1} shows the SPDs for the considered surface reconstructions of the CsI-T surfaces (SPDs of PbI$_2$-T reconstructed models are given in Fig.~S4 in SM). In principle, we need to plot the SPD in 3 dimensions (3D) because it depends on three chemical potentials:  $\Delta\mu_{\text{Cs}}^{}$\,, $\Delta\mu_{\text{Pb}}^{}$\,, and $\Delta\mu_{\text{I}}^{}$\,. Since such a 3D diagram is hard to visualize, we present two 2D slices instead, one $\Delta\mu_{\text{I}}^{}$/$\Delta\mu_{\text{Cs}}^{}$ slice at $\Delta\mu_{\mathrm{Pb}} = 0$ and one $\Delta\mu_{\text{I}}^{}$/$\Delta\mu_{\text{Pb}}^{}$ at $\Delta\mu_{\mathrm{Cs}}=0$. The left side of Fig.~\ref{spds_1} shows the SPDs at $\Delta\mu_{\text{Pb}} = 0$ and the right shows $\Delta\mu_{\text{Cs}} = 0$. Vertical panels show the $\upalpha$ and the $\upgamma$ phases, respectively. $\Delta\mu_{\text{I}}^{}$ is plotted on the vertical axis and the other chemical potential plotted on the horizontal axis. Colored regions and their labels indicate the most stable surface at that pair of chemical potentials. The white shaded region again depicts the growth limit for stable bulk $\text{CsPbI}_3^{}$\,, which serves as our reference to determine the most relevant surface models.

For the $\upalpha$ phase (Fig.~\ref{spds_1} upper panel), we find the following stable surface structures at some point in the phase diagram in the Pb-rich limit ($\Delta\mu_{\mathrm{Pb}}=0$):  v$_{4\text{Cs}}$\,, v$_{\text{CsI}}$\,, v$_{2\text{CsI}}$\,, v$_{\text{PbI}_2^{}}$\,, v$_{2\text{PbI}_2^{}}$\,, i$_{4\text{CsI}}$\,, i$_{\text{PbI}_2^{}}$\,,  i$_{2\mathrm{PbI}_2^{}}$\,, and the clean surface.  In the Cs-rich limit ($\Delta\mu_{\mathrm{Cs}}=0$), we instead find  v$_{2\text{PbI}_2^{}}$ and i$_{4\text{CsI}}$\,. The situation for the $\upgamma$ phase is very similar. The extent of some of the stability regions changes slightly from $\upalpha$ to $\upgamma$, and the v$_{\text{CsI}}$ reconstruction disappears from the phase diagram (Fig.~\ref{spds_1} lower panel). 

With the exception of v$_{4\text{Cs}}$\,, all the observed reconstructions are valence-neutral, i.e., with addition or removal of valence-neutral units such as CsI or PbI$_2$\,. Here, valence-neutral units refer to added or removed complexes that do not induce ``net charges`` on CsPbI$_3$ as a whole. The addition or removal of valence-neutral units is energetically more favorable than that of single atoms or non-valence-neutral complexes, because it does not introduce free charge carriers, as we will demonstrate in Sec.~\ref{elec_prop}. As expected, we observe Cs-deficient reconstructed models (v$_{4\text{Cs}}$\,, v$_{\text{CsI}}$\,, v$_{2\text{CsI}}$) for low $\Delta\mu_{\text{Cs}}^{}$ and Cs-rich ones (i$_{4\text{CsI}}$) at the high $\Delta\mu_{\text{Cs}}^{}$ region. A similar trend is observed for low and high Pb chemical potentials. A notable exception is the stability of i$_{4\text{CsI}}$ in the Pb-rich region in the upper right panel. Since the Cs chemical potential is at a maximum, the Cs-rich, i$_{4\text{CsI}}$ reconstruction dominates over Pb add-atom structures.

Of particular relevance to us are the surface reconstructions that intersect the bulk stability region (white region). These stable reconstructions that intersect the bulk region are the same for the $\upalpha$ and the $\upgamma$ phases. In addition to the clean CsI-T surface, we find only the valence-neutral surface reconstructions v$_{2\text{PbI}_2^{}}$\,, v$_{2\text{PbI}_2^{}}$\,, i$_{\text{PbI}_2^{}}$\,, i$_{\text{2PbI}_2^{}}$\,, and i$_{\text{4CsI}}$\,. It is noteworthy that although the clean surface occupies quite a broad stability region for $\Delta\mu_{\text{Pb}}=0$, it is only stable if the growth conditions are I-deficient and not at all in Cs-rich conditions. 

\subsubsection{Atomic structures of the most relevant surface reconstructions}\label{struct}

\begin{figure*}[!htp]
\includegraphics[width=\linewidth]{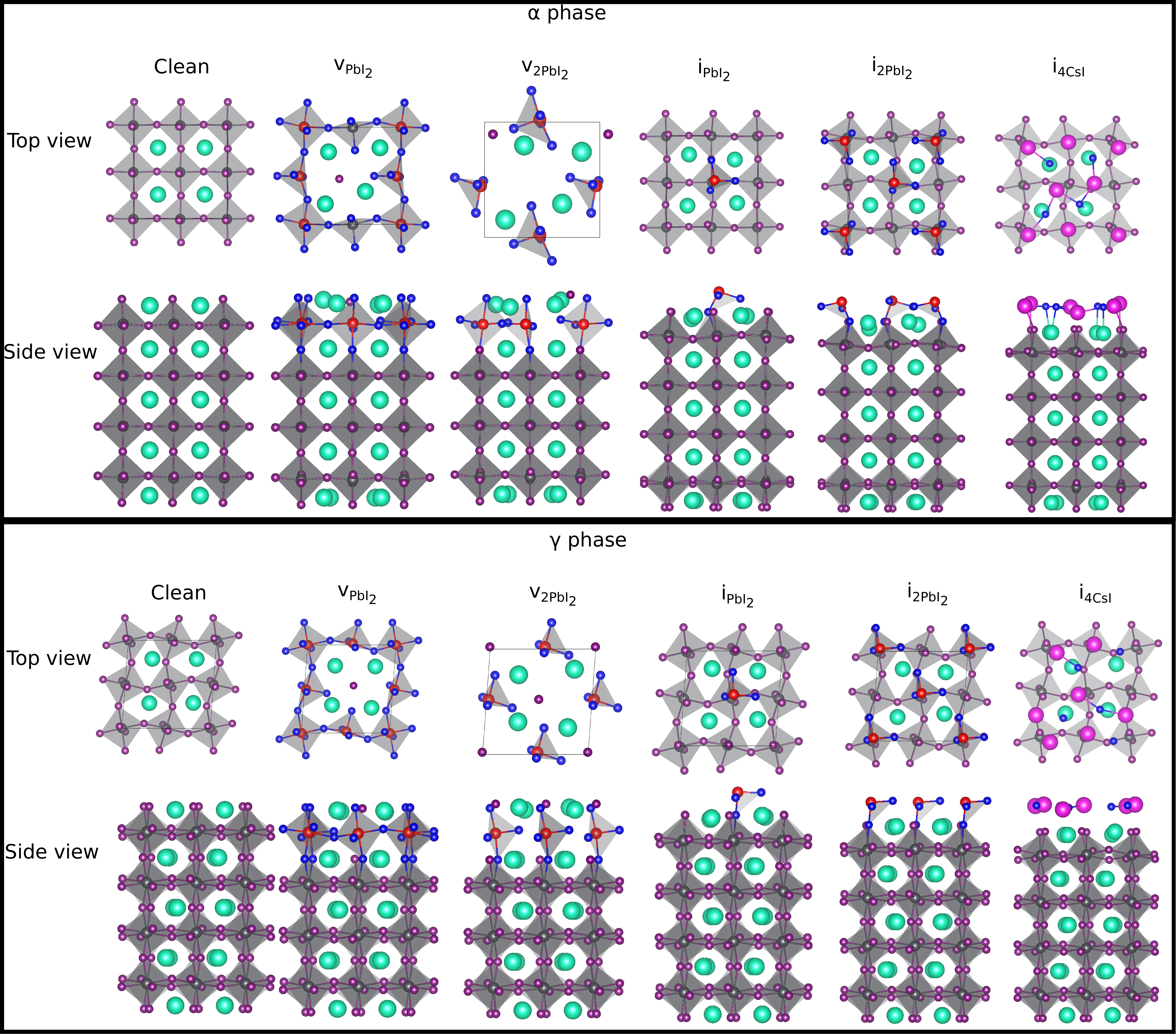}\caption{Atomic structures of the most relevant surface reconstructions for the $\upalpha$ phase (upper panel) and the $\upgamma$ phase (lower panel). The surface Pb and I atoms that exhibit pronounced displacements are highlighted by red and blue colors, respectively. The pink circles denote added Cs atoms.}
\label{surf_struct}
\end{figure*}

Figure~\ref{surf_struct} shows the relaxed geometries of the most relevant surface models (clean, v$_{\text{PbI}_2^{}}$\,,  v$_{2\text{PbI}_2^{}}$\,,  i$_{\text{PbI}_2^{}}$\,, i$_{2\text{PbI}_2^{}}$\,, and i$_{4\text{CsI}}$) for the $\upalpha$ and the $\upgamma$ phases. The remaining surface structures are shown in Fig.~\ref{csi}. The clean surface does not exhibit any significant deviations from the bulk atomic positions after relaxation. Hence, all changes in the reconstructed structures will be discussed with reference to the clean surface hereafter. 

In all reconstructions, we observe changes in the surface layer that translate into slight tilting of the surface octahedra. For instance, the surface octahedra in v$_{\text{PbI}_2^{}}$ for both phases tilt to account for the missing PbI$_2$ units. Similarly, the surface octahedra of i$_{\text{PbI}_2^{}}$ and i$_{2\text{PbI}_2^{}}$ tilt to accommodate the added PbI$_2$\,. In addition to the tilting, other slight changes in the atomic positions are observed. For example, the Cs-I bond lengths in the surfaces of i$_{4\text{CsI}}$ for both phases vary by $\sim0.1$~{\AA}. To highlight the changes in atomic positions, Cs, Pb, and I atoms of interest in Fig.~\ref{surf_struct} are depicted in pink, red and blue, respectively.

A more drastic change occurs for v$_{2\text{PbI}_2^{}}$ in the $\upalpha$ phase. Migration of I atoms within the surface layer leads to an asymmetric distribution of them, causing the formation of separate $\text{PbI}_5^{}$ and $\text{PbI}_4^{}$ polyhedra in the surface layer. This structure is similar to the findings of Haruyama et~al. for MAPbI$_3$ \cite{Haruyama14, Haruyama15, Haruyama16}, in which they observed the formation of $\text{PbI}_3^{}$-$\text{PbI}_5^{}$ polyhedra upon the removal of  ``one-half`` of the PbI$_2^{}$ units from the PbI$_2^{}$-T surfaces. Interestingly, we do not see the same atomic rearrangement for v$_{2\text{PbI}_2^{}}$ in the $\upgamma$ phase. Instead, we find a relatively symmetric I distribution and two isolated PbI$_4^{}$ polyhedra. The different behavior in the $\upalpha$ phase is likely due to the larger lattice constant, which results in a larger Cs--Cs distance and a weaker binding of I atoms. 

I migration is also observed in the i$_{2\text{I}}$ reconstruction of both phases and v$_{4\text{Cs}}$ of the $\upgamma$ phase (see Fig.~S1 in SM). In each case, two I atoms move close to each other such that their distance is close to that of an I$_2$ molecule \cite{NIST}. Specifically, the I-I distance in  v$_{4\text{Cs}}$ is reduced to $\sim 2.9$~{\AA}. Similarly, the I--I distance of the added I atoms in i$_{2\text{I}}$ reduces to  $\sim 2.9$~{\AA} in the $\upalpha$ phase and $\sim2.8$~{\AA} in the $\upgamma$ phase. These values are close to the experimental bond length ($\sim 2.67$~{\AA}) of the I$_2$ molecule in the gas phase \cite{NIST}, albeit a bit larger since the surface I atoms are bound to other surface atoms such as Cs and Pb, thus reducing the bond strength of I--I.

\subsection{Electronic properties of most relevant CsI-T surfaces}\label{elec_prop}\vspace{-0.5em}

In this section, we discuss the electronic properties of bulk CsPbI$_3$\,, the clean surfaces in the $\upalpha$ and the $\upgamma$ phases, as well as the relevant reconstructions reported in Figs.~\ref{spds_1} and \ref{surf_struct}.

\subsubsection{Electronic properties of the bulk and the clean CsI-T surface}\label{epb}

\begin{figure*}[!htp]
\includegraphics[width=\linewidth]{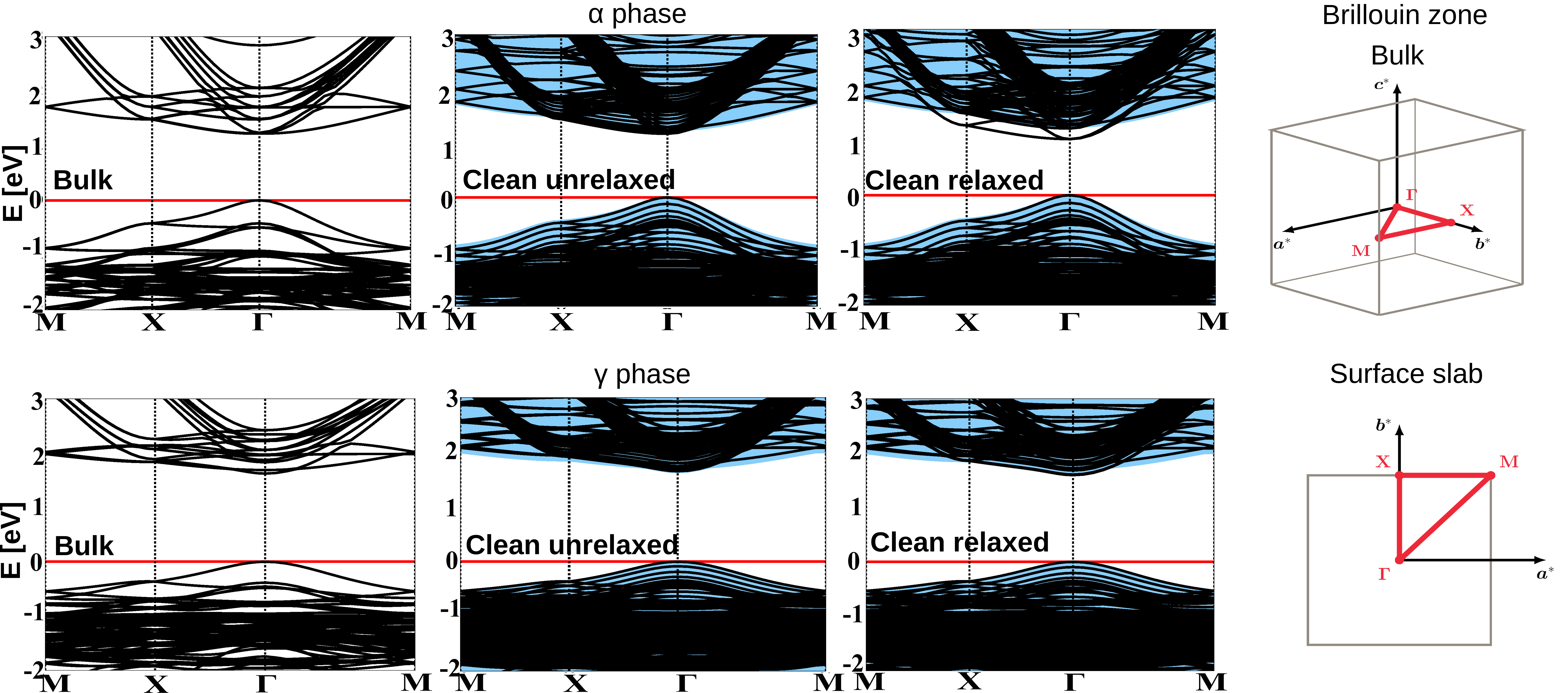}\vspace{-1.0em}
\caption  {Band structures of bulk CsPbI$_3$ and the clean surfaces (with and without geometry relaxation) in the $\upalpha$ (upper panel) and $\upgamma$ (lower panel) phases. Both bulk and surface band structures are calculated with a $2\times2$ in-plane supercell to share a common Brillouin zone and $k$-point path. As shown in the Brillouin zone (far right), in-plane $k$-point path for the bulk are the same as the surface. The VBM is set to 0 as marked by the red horizontal line. In the surface band structure plots, the projected bulk band structure is shown as blue shading.}
\label{bands}
\end{figure*}

Figure~\ref{bands} depicts the band structures of the bulk and the clean CsI-T surfaces of the $\upalpha$ and the $\upgamma$ phases. For the bulk of both phases, we adopt the high-symmetry $k$-point path in the Brillouin zone for a simple-cubic lattice of the $2\times2\times2$ supercell model for simplicity. In addition, we only show the band structure along M--X--$\Gamma$--M with $\text{M}=\big(\frac{1}{2},\frac{1}{2},0\big)$, $\text{X}=\big(0,\frac{1}{2},0\big)$, and $\Gamma=(0,0,0)$, i.e., within the $a^{\ast}b^{\ast}$ plane of the Brillouin zone (identical to the $ab=(001)$ plane in real space in our cases). Accordingly, we plot the band structure of the $2\times2$ surface unit cells of both phases along the same high-symmetry $k$-point path for an easy comparison. The valence band maximum (VBM) in all plots is set to zero. In the band structure plots of the surface models, the projected bulk band structure is included  as a blue-shaded background to help identify possible surface states \cite{Inglesfield82, Speer09}.

\begin{figure}[!htp]
\includegraphics[width=\linewidth]{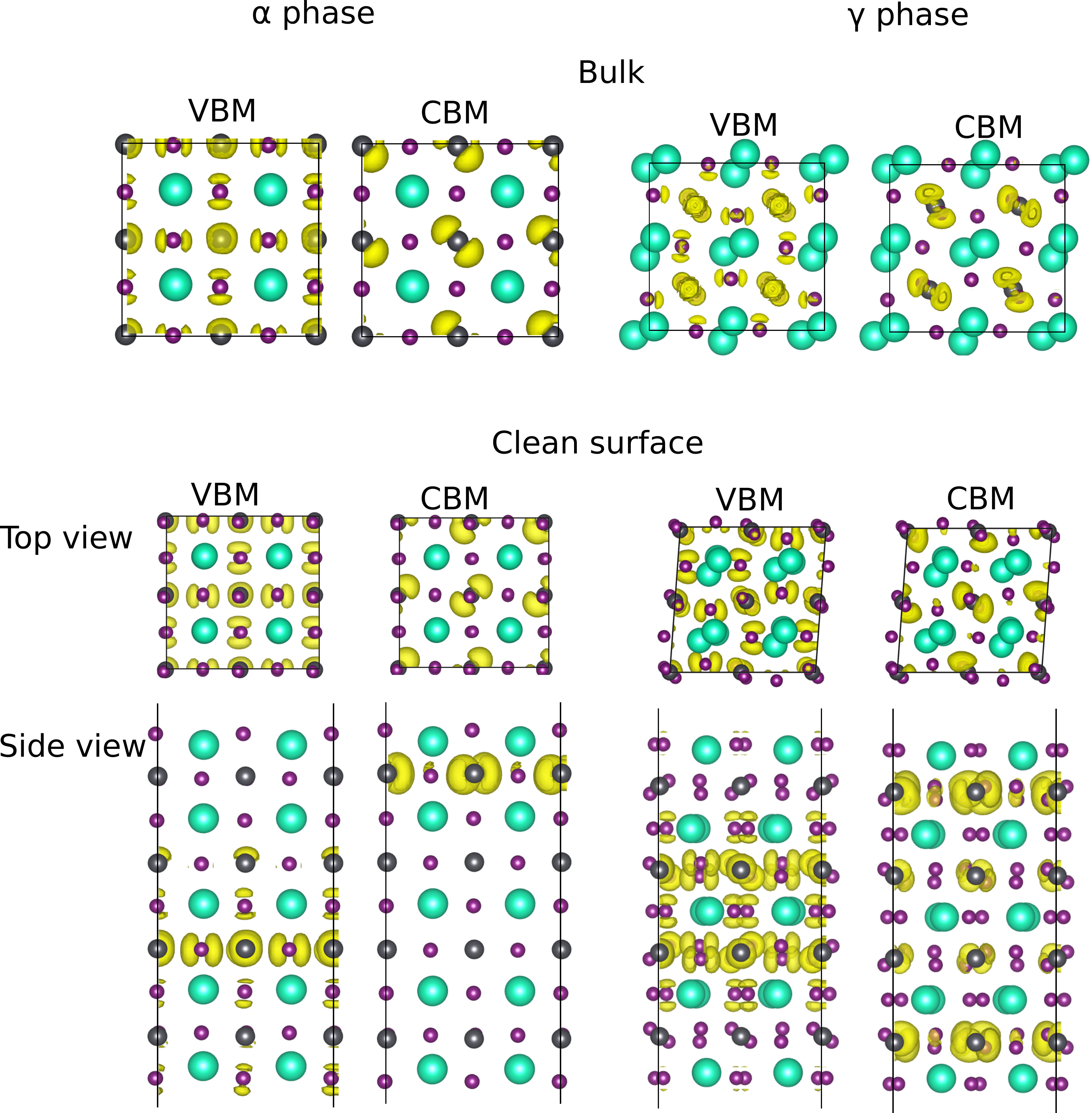}
\caption{Charge distribution of the VBM and CBM of bulk CsPbI$_3$ and the CsI-T clean surfaces for (a) the $\upalpha$ and (b) the $\upgamma$ phase.} \label{cube} 
\end{figure}

In both phases, bulk CsPbI$_3$ exhibits a direct band gap at the $\Gamma$ point. The charge densities (shown in Fig.~\ref{cube}) reveal that the VBM of CsPbI$_3$ in both phases is dominated by I-5p orbitals with a noticeable contribution from the Pb-6s orbitals, which gives rise to the well known antibonding character \cite{Huang18, Haruyama14, Haruyama16, Wei17}. The conduction band maximum (CBM) mainly consists of Pb-6p orbitals.

Next, we investigate if the CsI termination introduces surface states. The middle panels of  Fig.~\ref{bands} show the band structure of the two unrelaxed CsI-T models. The bands of the supercell coincide with the projected bulk band structure, which indicates that no surface states appear. However, upon relaxation, the bottom of the conduction band is pulled into the bulk band gap for the $\upalpha$ phase but not for the $\upgamma$ phase, as can be seen in the right most band structure panels in Fig.~\ref{bands}. The clean surface of the $\upalpha$ phase therefore exhibits a surface state that derives from the CsPbI$_3$ conduction band. This is further evidenced in Fig.~\ref{cube}, which shows that the lowest conduction band resides at the surface and has Pb-6p character. In contrast, the corresponding state in the $\upgamma$ phase is quite clearly a bulk state and not a surface state.

\subsubsection{Electronic properties of most relevant reconstructed CsI-T surface models}

\begin{figure*}[!ht]
\includegraphics[width=\linewidth]{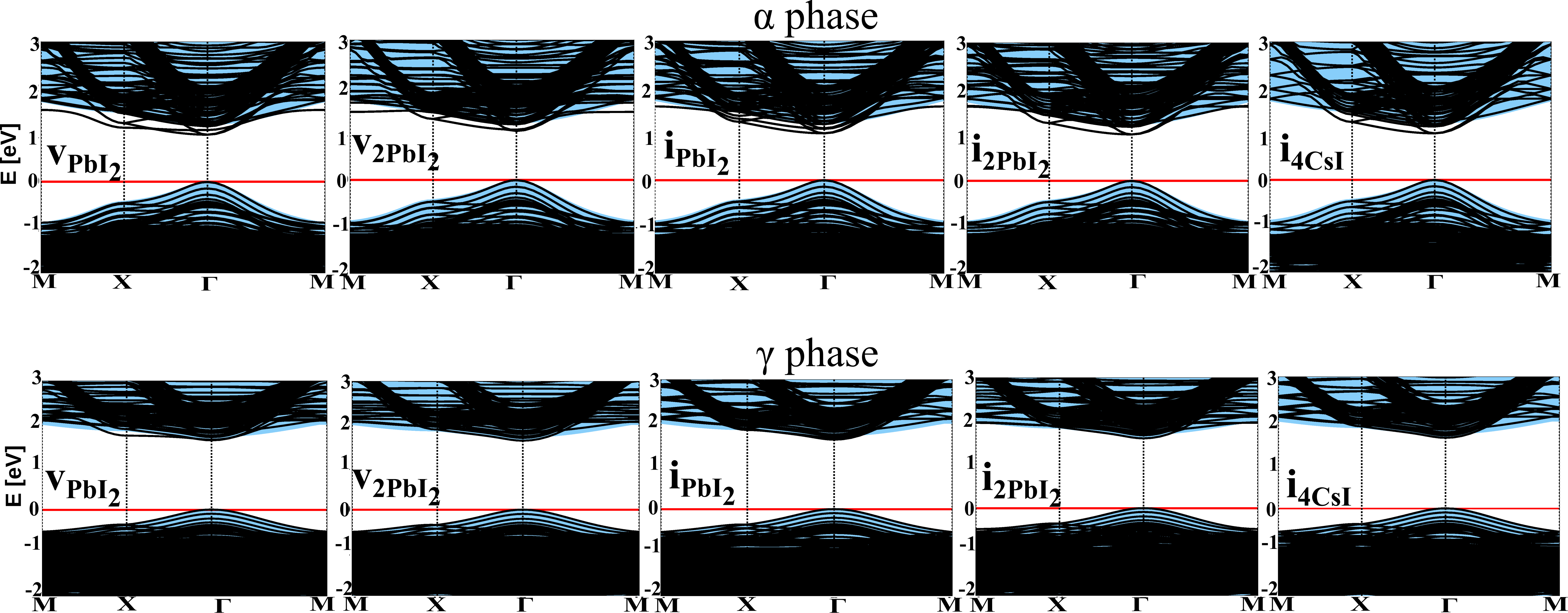}
\caption{Band structures of the most relevant reconstructed CsI-T surface models in the $\upalpha$ and the $\upgamma$ phases. Legends follow Fig.\ref{bands}.} \label{bands_1}
\end{figure*}

Figure~\ref{bands_1} shows the band structures of the most relevant surface models observed in Fig.~\ref{spds_1}, i.e., v$_{\mathrm{PbI}_2^{}}$\,, v$_{2\mathrm{PbI}_2^{}}$\,, i$_{4\mathrm{CsI}}$\,, i$_{\mathrm{PbI}_2^{}}$ and i$_{2\mathrm{PbI}_2^{}}$\,. Similar to Fig.~\ref{bands}, the bulk band structure is included as the background for comparison.

Figure~\ref{bands_1} displays a similar pattern as  Fig.~\ref{bands}, i.e., the most notable changes in the band structure of the surface models appear for the $\upalpha$ phase near the bottom of the conduction band. For neither phase do we observe perturbations of the VBM region. Further inspection of the charge distributions of the valence-band-edge states shown in Fig.~S2 in SM confirms that the VBM retains its bulk character for all relevant surface reconstructions.

For the reconstructed surface models of the $\upalpha$ phase, we observe the same surface states as in the clean-surface model (Fig.~\ref{bands}). In addition, flat bands appear near or below the conduction band edge that are most notable around the M-points of the band structure. The only exception is  i$_{4\text{CsI}}$\,, for which the surface band structure strongly resembles that of the clean CsI-T surface. The flat bands are especially pronounced in the v$_{\text{PbI}_2}$ and v$_{2\text{PbI}_2}$ models. Correspondingly, the states near the conduction-band edge of these two surface reconstructions exhibit a more localized character than for the clean surface (compare Fig.~S2 upper panel in SM with Fig.~\ref{cube}). For these surface models, the reconstruction therefore introduces additional surface states to the ones of the clean surface.

\section{Discussion}

Our results offer guidance for growing favorable CsPbI$_3$ surfaces. By favorable, we here imply surface reconstructions that have a bulk-like band structure and no additional states in the band gap or perturbations of the band edges that might adversely affect the transport properties. Our analysis of the previous section suggests that the $\upgamma$ phase of CsPbI$_3$ is generally more suited for this purpose, as all of its stable surface reconstructions are free of band edge perturbations.

For the $\upalpha$ phase, the objective would be to avoid both PbI$_2$ deficient (v$_{\mathrm{PbI}_2}$ and v$_{\mathrm{2PbI}_2}$) and rich (i$_{\mathrm{PbI}_2}$ and i$_{\mathrm{2PbI}_2}$) reconstructions. Fortunately, the clean surface is stable across a wide range of the bulk stability region, as our surface phase diagram analysis shows. For Cs rich growth conditions, the i$_{\mathrm{4CsI}}$ phase dominates the phase diagram. This phase provides a good alternative to the clean surface since its band structure resembles that of the clean surface closely.

\section{Conclusions}\label{conclusion}

In summary, we have studied the surface atomic and electronic structure of  CsPbI$_3^{}$ from first principles. For both the $\upalpha$ (cubic) and the $\upgamma$ (orthorhombic) phases, we have considered the clean-surface models and a series of surface reconstructions. Surface phase diagram analysis indicates that the CsI-terminated $(001)$ surface is more stable within a large range of allowed chemical potentials for both phases. In addition, several CsI and PbI$_2$ rich and deficient surface reconstructions are stable. These surface reconstructions do not induce deep energy levels in the band gap. Nevertheless, the removal of PbI$_2^{}$ units in the CsI-terminated $\upalpha$-CsPbI$_3^{}$ surface has noticeable effects on the material's electronic structure, especially close to the conduction-band edge. Combining our surface-phase diagram and electronic structure analysis allows us to recommend growth regimes for CsPbI$_3^{}$ surfaces with favorable transport properties. Our work highlights the complexity of CsPbI$_3^{}$ surfaces and provides avenues for future surface science and interface studies.

\section{Supplementary Material}
See Supplementary Material for the charge density plots of all relevant surface models with CsI-termination, surface phase diagrams of PbI$_2$-terminated models, crystal structures of studied surface models in both CsI-T and PbI$_2$-T (that are not included in the main text) and the formation energies of all surface models.  

\section*{Acknowledgments}
We acknowledge the computing resources by the CSC-IT Center for Science, the Aalto Science-IT project, and Xi’an Jiaotong University’s HPC Platform. An award of computer time was provided by the Innovative and Novel Computational Impact on Theory and Experiment (INCITE) program. This research used resources of the Argonne Leadership Computing Facility, which is a DOE Office of Science User Facility supported under Contract DE-AC02-06CH11357. We further acknowledge funding  from the European Union’s Horizon 2020 research and innovation programme under grant agreement No 676580 (The Novel Materials Discovery (NOMAD) Laboratory), the V\"ais\"al\"a Foundation as well as the Academy of Finland through its Centres of Excellence Programme (284621), its Key Project Funding scheme (305632) and project 316347. 

\section{Aip publishing data sharing policy} 
The data that supports the findings of this study will be openly available in NOvel Materials Discovery (NOMAD) repository at \cite{NOMAD}.

\newpage
\bibliographystyle{apsrev}
\bibliography{aipsamp}

\begin{thebibliography}{77}
\expandafter\ifx\csname natexlab\endcsname\relax\def\natexlab#1{#1}\fi
\expandafter\ifx\csname bibnamefont\endcsname\relax
  \def\bibnamefont#1{#1}\fi
\expandafter\ifx\csname bibfnamefont\endcsname\relax
  \def\bibfnamefont#1{#1}\fi
\expandafter\ifx\csname citenamefont\endcsname\relax
  \def\citenamefont#1{#1}\fi
\expandafter\ifx\csname url\endcsname\relax
  \def\url#1{\texttt{#1}}\fi
\expandafter\ifx\csname urlprefix\endcsname\relax\def\urlprefix{URL }\fi
\providecommand{\bibinfo}[2]{#2}
\providecommand{\eprint}[2][]{\url{#2}}

\bibitem[{NRE(2020)}]{NRELchart19}
\emph{\bibinfo{title}{\htmladdnormallink{\color{blue}{https://www.nrel.gov/pv/assets/pdfs/best-research-cell-efficiencies.20200311.pdf}}{https://www.nrel.gov/pv/assets/pdfs/best-research-cell-efficiencies.20200311.pdf}
  \\}} (\bibinfo{publisher}{National Renewable Energy Laboratory: Best
  research-cell efficiencies}, \bibinfo{year}{2020}).

\bibitem[{\citenamefont{Kim et~al.}(2012)\citenamefont{Kim, Lee, Im, Lee,
  Moehl, Marchioro, Moon, Humphry-Baker, Yum, Moser et~al.}}]{kim}
\bibinfo{author}{\bibfnamefont{H.-S.} \bibnamefont{Kim}},
  \bibinfo{author}{\bibfnamefont{C.-R.} \bibnamefont{Lee}},
  \bibinfo{author}{\bibfnamefont{J.-H.} \bibnamefont{Im}},
  \bibinfo{author}{\bibfnamefont{K.-B.} \bibnamefont{Lee}},
  \bibinfo{author}{\bibfnamefont{T.}~\bibnamefont{Moehl}},
  \bibinfo{author}{\bibfnamefont{A.}~\bibnamefont{Marchioro}},
  \bibinfo{author}{\bibfnamefont{S.-J.} \bibnamefont{Moon}},
  \bibinfo{author}{\bibfnamefont{R.}~\bibnamefont{Humphry-Baker}},
  \bibinfo{author}{\bibfnamefont{J.-H.} \bibnamefont{Yum}},
  \bibinfo{author}{\bibfnamefont{J.~E.} \bibnamefont{Moser}},
  \bibnamefont{et~al.}, \bibinfo{journal}{Sci. Rep.}
  \textbf{\bibinfo{volume}{2}}, \bibinfo{pages}{591} (\bibinfo{year}{2012}).

\bibitem[{\citenamefont{Lee et~al.}(2012)\citenamefont{Lee, Teuscher, Miyasaka,
  Murakami, and Snaith}}]{lee}
\bibinfo{author}{\bibfnamefont{M.~M.} \bibnamefont{Lee}},
  \bibinfo{author}{\bibfnamefont{J.}~\bibnamefont{Teuscher}},
  \bibinfo{author}{\bibfnamefont{T.}~\bibnamefont{Miyasaka}},
  \bibinfo{author}{\bibfnamefont{T.~N.} \bibnamefont{Murakami}},
  \bibnamefont{and} \bibinfo{author}{\bibfnamefont{H.}~\bibnamefont{Snaith}},
  \bibinfo{journal}{Science} \textbf{\bibinfo{volume}{338}},
  \bibinfo{pages}{643} (\bibinfo{year}{2012}).

\bibitem[{\citenamefont{Huang and Lambrecht}(2014)}]{Huang14}
\bibinfo{author}{\bibfnamefont{L.-Y.} \bibnamefont{Huang}} \bibnamefont{and}
  \bibinfo{author}{\bibfnamefont{W.~R.~L.} \bibnamefont{Lambrecht}},
  \bibinfo{journal}{Phys. Rev. B} \textbf{\bibinfo{volume}{90}},
  \bibinfo{pages}{195201} (\bibinfo{year}{2014}).

\bibitem[{\citenamefont{Lin et~al.}(2015)\citenamefont{Lin, Armin, Lyons, Burn,
  and Meredith}}]{Lin15}
\bibinfo{author}{\bibfnamefont{Q.}~\bibnamefont{Lin}},
  \bibinfo{author}{\bibfnamefont{A.}~\bibnamefont{Armin}},
  \bibinfo{author}{\bibfnamefont{D.~M.} \bibnamefont{Lyons}},
  \bibinfo{author}{\bibfnamefont{P.~L.} \bibnamefont{Burn}}, \bibnamefont{and}
  \bibinfo{author}{\bibfnamefont{P.}~\bibnamefont{Meredith}},
  \bibinfo{journal}{Advanced Materials} pp. \bibinfo{pages}{2060--2064}
  (\bibinfo{year}{2015}).

\bibitem[{\citenamefont{Cho et~al.}(2015)\citenamefont{Cho, Jeong, Park, Kim,
  Wolf, Lee, Heo, Sadhanala, Myoung, Yoo et~al.}}]{Cho15}
\bibinfo{author}{\bibfnamefont{H.}~\bibnamefont{Cho}},
  \bibinfo{author}{\bibfnamefont{S.-H.} \bibnamefont{Jeong}},
  \bibinfo{author}{\bibfnamefont{M.-H.} \bibnamefont{Park}},
  \bibinfo{author}{\bibfnamefont{Y.-H.} \bibnamefont{Kim}},
  \bibinfo{author}{\bibfnamefont{C.}~\bibnamefont{Wolf}},
  \bibinfo{author}{\bibfnamefont{C.-L.} \bibnamefont{Lee}},
  \bibinfo{author}{\bibfnamefont{J.~H.} \bibnamefont{Heo}},
  \bibinfo{author}{\bibfnamefont{A.}~\bibnamefont{Sadhanala}},
  \bibinfo{author}{\bibfnamefont{N.}~\bibnamefont{Myoung}},
  \bibinfo{author}{\bibfnamefont{S.}~\bibnamefont{Yoo}}, \bibnamefont{et~al.},
  \bibinfo{journal}{Science} \textbf{\bibinfo{volume}{350}}
  (\bibinfo{year}{2015}).

\bibitem[{\citenamefont{Snaith}(2013)}]{snaith}
\bibinfo{author}{\bibfnamefont{H.~J.} \bibnamefont{Snaith}},
  \bibinfo{journal}{J. Phys. Chem. Lett.} \textbf{\bibinfo{volume}{4}},
  \bibinfo{pages}{3623} (\bibinfo{year}{2013}).

\bibitem[{\citenamefont{Green et~al.}(2014)\citenamefont{Green, Ho-Baille, and
  Snaith}}]{green}
\bibinfo{author}{\bibfnamefont{M.~A.} \bibnamefont{Green}},
  \bibinfo{author}{\bibfnamefont{A.}~\bibnamefont{Ho-Baille}},
  \bibnamefont{and} \bibinfo{author}{\bibfnamefont{H.~J.}
  \bibnamefont{Snaith}}, \bibinfo{journal}{Nat. Photon.}
  \textbf{\bibinfo{volume}{8}}, \bibinfo{pages}{506} (\bibinfo{year}{2014}).

\bibitem[{\citenamefont{Stranks et~al.}(2013)\citenamefont{Stranks, Eperon,
  Grancini, Menelaou, Alcocer, Leijtens, Herz, Petrozza, and Snaith}}]{stranks}
\bibinfo{author}{\bibfnamefont{S.~D.} \bibnamefont{Stranks}},
  \bibinfo{author}{\bibfnamefont{G.~E.} \bibnamefont{Eperon}},
  \bibinfo{author}{\bibfnamefont{G.}~\bibnamefont{Grancini}},
  \bibinfo{author}{\bibfnamefont{C.}~\bibnamefont{Menelaou}},
  \bibinfo{author}{\bibfnamefont{M.~J.~P.} \bibnamefont{Alcocer}},
  \bibinfo{author}{\bibfnamefont{T.}~\bibnamefont{Leijtens}},
  \bibinfo{author}{\bibfnamefont{L.~M.} \bibnamefont{Herz}},
  \bibinfo{author}{\bibfnamefont{A.}~\bibnamefont{Petrozza}}, \bibnamefont{and}
  \bibinfo{author}{\bibfnamefont{H.~J.} \bibnamefont{Snaith}},
  \bibinfo{journal}{Science} \textbf{\bibinfo{volume}{342}},
  \bibinfo{pages}{341} (\bibinfo{year}{2013}).

\bibitem[{\citenamefont{Xing et~al.}(2013)\citenamefont{Xing, Mathews, Sun,
  Lim, Lam, Gr\"atzel, Mhaisalkar, and Sum}}]{Xing13}
\bibinfo{author}{\bibfnamefont{G.}~\bibnamefont{Xing}},
  \bibinfo{author}{\bibfnamefont{N.}~\bibnamefont{Mathews}},
  \bibinfo{author}{\bibfnamefont{S.}~\bibnamefont{Sun}},
  \bibinfo{author}{\bibfnamefont{S.~S.} \bibnamefont{Lim}},
  \bibinfo{author}{\bibfnamefont{Y.~M.} \bibnamefont{Lam}},
  \bibinfo{author}{\bibfnamefont{M.}~\bibnamefont{Gr\"atzel}},
  \bibinfo{author}{\bibfnamefont{S.}~\bibnamefont{Mhaisalkar}},
  \bibnamefont{and} \bibinfo{author}{\bibfnamefont{T.~C.} \bibnamefont{Sum}},
  \bibinfo{journal}{Science} \textbf{\bibinfo{volume}{342}},
  \bibinfo{pages}{344} (\bibinfo{year}{2013}).

\bibitem[{\citenamefont{Eperon et~al.}(2015)\citenamefont{Eperon, Patern{\`o},
  Sutton, Zampetti, Haghighirad, Cacialli, and Snaith}}]{Eperon15}
\bibinfo{author}{\bibfnamefont{G.~E.} \bibnamefont{Eperon}},
  \bibinfo{author}{\bibfnamefont{G.~M.} \bibnamefont{Patern{\`o}}},
  \bibinfo{author}{\bibfnamefont{R.~J.} \bibnamefont{Sutton}},
  \bibinfo{author}{\bibfnamefont{A.}~\bibnamefont{Zampetti}},
  \bibinfo{author}{\bibfnamefont{A.~A.} \bibnamefont{Haghighirad}},
  \bibinfo{author}{\bibfnamefont{F.}~\bibnamefont{Cacialli}}, \bibnamefont{and}
  \bibinfo{author}{\bibfnamefont{H.~J.} \bibnamefont{Snaith}},
  \bibinfo{journal}{J. Mater. Chem. A} \textbf{\bibinfo{volume}{3}},
  \bibinfo{pages}{19688} (\bibinfo{year}{2015}).

\bibitem[{\citenamefont{Troughton et~al.}(2017)\citenamefont{Troughton, Hooper,
  and Watson}}]{troughton}
\bibinfo{author}{\bibfnamefont{J.}~\bibnamefont{Troughton}},
  \bibinfo{author}{\bibfnamefont{K.}~\bibnamefont{Hooper}}, \bibnamefont{and}
  \bibinfo{author}{\bibfnamefont{T.~M.} \bibnamefont{Watson}},
  \bibinfo{journal}{Nano Energy} \textbf{\bibinfo{volume}{39}},
  \bibinfo{pages}{60} (\bibinfo{year}{2017}).

\bibitem[{\citenamefont{Niu et~al.}(2014)\citenamefont{Niu, Li, Meng, Wang,
  Dong, and Qiu}}]{NiuG14}
\bibinfo{author}{\bibfnamefont{G.}~\bibnamefont{Niu}},
  \bibinfo{author}{\bibfnamefont{W.}~\bibnamefont{Li}},
  \bibinfo{author}{\bibfnamefont{F.}~\bibnamefont{Meng}},
  \bibinfo{author}{\bibfnamefont{L.}~\bibnamefont{Wang}},
  \bibinfo{author}{\bibfnamefont{H.}~\bibnamefont{Dong}}, \bibnamefont{and}
  \bibinfo{author}{\bibfnamefont{Y.}~\bibnamefont{Qiu}}, \bibinfo{journal}{J.
  Mater. Chem. A} \textbf{\bibinfo{volume}{2}}, \bibinfo{pages}{705}
  (\bibinfo{year}{2014}).

\bibitem[{\citenamefont{Niu et~al.}(2015)\citenamefont{Niu, Guo, and
  Wang}}]{NiuG15}
\bibinfo{author}{\bibfnamefont{G.}~\bibnamefont{Niu}},
  \bibinfo{author}{\bibfnamefont{X.}~\bibnamefont{Guo}}, \bibnamefont{and}
  \bibinfo{author}{\bibfnamefont{L.}~\bibnamefont{Wang}}, \bibinfo{journal}{J.
  Mater. Chem. A} \textbf{\bibinfo{volume}{3}}, \bibinfo{pages}{8970}
  (\bibinfo{year}{2015}).

\bibitem[{\citenamefont{Huang et~al.}(2017)\citenamefont{Huang, Tan, Lund, and
  Zhou}}]{HuangJ17}
\bibinfo{author}{\bibfnamefont{J.}~\bibnamefont{Huang}},
  \bibinfo{author}{\bibfnamefont{S.}~\bibnamefont{Tan}},
  \bibinfo{author}{\bibfnamefont{P.~D.} \bibnamefont{Lund}}, \bibnamefont{and}
  \bibinfo{author}{\bibfnamefont{H.}~\bibnamefont{Zhou}},
  \bibinfo{journal}{Energy Environ. Sci.} \textbf{\bibinfo{volume}{10}},
  \bibinfo{pages}{2284} (\bibinfo{year}{2017}).

\bibitem[{\citenamefont{Kim et~al.}(2017)\citenamefont{Kim, Jang, Yoon, Jeong,
  Park, Walker, Jeon, Jo, Yoon, Kim et~al.}}]{KimGH17}
\bibinfo{author}{\bibfnamefont{G.-H.} \bibnamefont{Kim}},
  \bibinfo{author}{\bibfnamefont{H.}~\bibnamefont{Jang}},
  \bibinfo{author}{\bibfnamefont{Y.~J.} \bibnamefont{Yoon}},
  \bibinfo{author}{\bibfnamefont{J.}~\bibnamefont{Jeong}},
  \bibinfo{author}{\bibfnamefont{S.~Y.} \bibnamefont{Park}},
  \bibinfo{author}{\bibfnamefont{B.}~\bibnamefont{Walker}},
  \bibinfo{author}{\bibfnamefont{I.-Y.} \bibnamefont{Jeon}},
  \bibinfo{author}{\bibfnamefont{Y.}~\bibnamefont{Jo}},
  \bibinfo{author}{\bibfnamefont{H.}~\bibnamefont{Yoon}},
  \bibinfo{author}{\bibfnamefont{M.}~\bibnamefont{Kim}}, \bibnamefont{et~al.},
  \bibinfo{journal}{Nano Lett.} \textbf{\bibinfo{volume}{17}},
  \bibinfo{pages}{6385} (\bibinfo{year}{2017}).

\bibitem[{\citenamefont{Mesquita et~al.}(2018)\citenamefont{Mesquita, Andrade,
  and Mendes}}]{Mesquita18}
\bibinfo{author}{\bibfnamefont{I.}~\bibnamefont{Mesquita}},
  \bibinfo{author}{\bibfnamefont{L.}~\bibnamefont{Andrade}}, \bibnamefont{and}
  \bibinfo{author}{\bibfnamefont{A.}~\bibnamefont{Mendes}},
  \bibinfo{journal}{Renew. Sustain. Energy Rev.} \textbf{\bibinfo{volume}{82}},
  \bibinfo{pages}{2471} (\bibinfo{year}{2018}).

\bibitem[{\citenamefont{Ciccioli and Latini}(2018)}]{Ciccioli18}
\bibinfo{author}{\bibfnamefont{A.}~\bibnamefont{Ciccioli}} \bibnamefont{and}
  \bibinfo{author}{\bibfnamefont{A.}~\bibnamefont{Latini}},
  \bibinfo{journal}{J. Phys. Chem. Lett.} \textbf{\bibinfo{volume}{9}},
  \bibinfo{pages}{3756} (\bibinfo{year}{2018}).

\bibitem[{\citenamefont{Li et~al.}(2018)\citenamefont{Li, Yuan, Ling, Zhang,
  Yang, Cheung, Ho, Gao, and Ma}}]{LiF18}
\bibinfo{author}{\bibfnamefont{F.}~\bibnamefont{Li}},
  \bibinfo{author}{\bibfnamefont{J.}~\bibnamefont{Yuan}},
  \bibinfo{author}{\bibfnamefont{X.}~\bibnamefont{Ling}},
  \bibinfo{author}{\bibfnamefont{Y.}~\bibnamefont{Zhang}},
  \bibinfo{author}{\bibfnamefont{Y.}~\bibnamefont{Yang}},
  \bibinfo{author}{\bibfnamefont{S.~H.} \bibnamefont{Cheung}},
  \bibinfo{author}{\bibfnamefont{C.~H.~Y.} \bibnamefont{Ho}},
  \bibinfo{author}{\bibfnamefont{X.}~\bibnamefont{Gao}}, \bibnamefont{and}
  \bibinfo{author}{\bibfnamefont{W.}~\bibnamefont{Ma}}, \bibinfo{journal}{Adv.
  Funct. Mater.} \textbf{\bibinfo{volume}{18}}, \bibinfo{pages}{1706377}
  (\bibinfo{year}{2018}).

\bibitem[{\citenamefont{Schmidt et~al.}(2014)\citenamefont{Schmidt,
  Perteg{\'a}, Gonz{\'a}lez-Carrero, Malinkiewicz, Agouram, Espallargas,
  Bolink, Galian, and P{\'e}rez-Prieto}}]{SchmidtL14}
\bibinfo{author}{\bibfnamefont{L.~C.} \bibnamefont{Schmidt}},
  \bibinfo{author}{\bibfnamefont{A.}~\bibnamefont{Perteg{\'a}}},
  \bibinfo{author}{\bibfnamefont{S.}~\bibnamefont{Gonz{\'a}lez-Carrero}},
  \bibinfo{author}{\bibfnamefont{O.}~\bibnamefont{Malinkiewicz}},
  \bibinfo{author}{\bibfnamefont{S.}~\bibnamefont{Agouram}},
  \bibinfo{author}{\bibfnamefont{G.~M.} \bibnamefont{Espallargas}},
  \bibinfo{author}{\bibfnamefont{H.~J.} \bibnamefont{Bolink}},
  \bibinfo{author}{\bibfnamefont{R.~E.} \bibnamefont{Galian}},
  \bibnamefont{and}
  \bibinfo{author}{\bibfnamefont{J.}~\bibnamefont{P{\'e}rez-Prieto}},
  \bibinfo{journal}{J. Am. Chem. Soc.} \textbf{\bibinfo{volume}{136}},
  \bibinfo{pages}{850} (\bibinfo{year}{2014}).

\bibitem[{\citenamefont{Gonz{\'a}lez-Carrero
  et~al.}(2015)\citenamefont{Gonz{\'a}lez-Carrero, Galian, and
  P{\'e}rez-Prieto}}]{SoranyelG15}
\bibinfo{author}{\bibfnamefont{S.}~\bibnamefont{Gonz{\'a}lez-Carrero}},
  \bibinfo{author}{\bibfnamefont{R.~E.} \bibnamefont{Galian}},
  \bibnamefont{and}
  \bibinfo{author}{\bibfnamefont{J.}~\bibnamefont{P{\'e}rez-Prieto}},
  \bibinfo{journal}{J. Mater. Chem. A} \textbf{\bibinfo{volume}{3}},
  \bibinfo{pages}{9187} (\bibinfo{year}{2015}).

\bibitem[{\citenamefont{Dong et~al.}(2019)\citenamefont{Dong, Xi, Zuo, Li,
  Yang, Wang, Yu, Ma, Ran, Gao et~al.}}]{Dong19}
\bibinfo{author}{\bibfnamefont{H.}~\bibnamefont{Dong}},
  \bibinfo{author}{\bibfnamefont{J.}~\bibnamefont{Xi}},
  \bibinfo{author}{\bibfnamefont{L.}~\bibnamefont{Zuo}},
  \bibinfo{author}{\bibfnamefont{J.}~\bibnamefont{Li}},
  \bibinfo{author}{\bibfnamefont{Y.}~\bibnamefont{Yang}},
  \bibinfo{author}{\bibfnamefont{D.}~\bibnamefont{Wang}},
  \bibinfo{author}{\bibfnamefont{Y.}~\bibnamefont{Yu}},
  \bibinfo{author}{\bibfnamefont{L.}~\bibnamefont{Ma}},
  \bibinfo{author}{\bibfnamefont{C.}~\bibnamefont{Ran}},
  \bibinfo{author}{\bibfnamefont{W.}~\bibnamefont{Gao}}, \bibnamefont{et~al.},
  \bibinfo{journal}{Adv. Funct. Mater.} \textbf{\bibinfo{volume}{29}},
  \bibinfo{pages}{1808119} (\bibinfo{year}{2019}).

\bibitem[{\citenamefont{Quan et~al.}(2016)\citenamefont{Quan, Yuan, Comin,
  Voznyy, Beauregard, Hoogland, Buin, Kirmani, Zhao, Amassian et~al.}}]{Quan16}
\bibinfo{author}{\bibfnamefont{L.~N.} \bibnamefont{Quan}},
  \bibinfo{author}{\bibfnamefont{M.}~\bibnamefont{Yuan}},
  \bibinfo{author}{\bibfnamefont{R.}~\bibnamefont{Comin}},
  \bibinfo{author}{\bibfnamefont{O.}~\bibnamefont{Voznyy}},
  \bibinfo{author}{\bibfnamefont{E.~M.} \bibnamefont{Beauregard}},
  \bibinfo{author}{\bibfnamefont{S.}~\bibnamefont{Hoogland}},
  \bibinfo{author}{\bibfnamefont{A.}~\bibnamefont{Buin}},
  \bibinfo{author}{\bibfnamefont{A.~R.} \bibnamefont{Kirmani}},
  \bibinfo{author}{\bibfnamefont{K.}~\bibnamefont{Zhao}},
  \bibinfo{author}{\bibfnamefont{A.}~\bibnamefont{Amassian}},
  \bibnamefont{et~al.}, \bibinfo{journal}{J. Am. Chem. Soc.}
  \textbf{\bibinfo{volume}{138}}, \bibinfo{pages}{2649} (\bibinfo{year}{2016}).

\bibitem[{\citenamefont{Dou}(2017)}]{Dou17}
\bibinfo{author}{\bibfnamefont{L.}~\bibnamefont{Dou}}, \bibinfo{journal}{J.
  Mater. Chem. C} \textbf{\bibinfo{volume}{5}}, \bibinfo{pages}{11165}
  (\bibinfo{year}{2017}).

\bibitem[{\citenamefont{Ran et~al.}(2018)\citenamefont{Ran, Xi, Gao, Yuan, Lei,
  Jiao, Hou, and Wu}}]{Ran18a}
\bibinfo{author}{\bibfnamefont{C.}~\bibnamefont{Ran}},
  \bibinfo{author}{\bibfnamefont{J.}~\bibnamefont{Xi}},
  \bibinfo{author}{\bibfnamefont{W.}~\bibnamefont{Gao}},
  \bibinfo{author}{\bibfnamefont{F.}~\bibnamefont{Yuan}},
  \bibinfo{author}{\bibfnamefont{T.}~\bibnamefont{Lei}},
  \bibinfo{author}{\bibfnamefont{B.}~\bibnamefont{Jiao}},
  \bibinfo{author}{\bibfnamefont{X.}~\bibnamefont{Hou}}, \bibnamefont{and}
  \bibinfo{author}{\bibfnamefont{Z.}~\bibnamefont{Wu}}, \bibinfo{journal}{ACS
  Energy Lett.} \textbf{\bibinfo{volume}{3}}, \bibinfo{pages}{713}
  (\bibinfo{year}{2018}).

\bibitem[{\citenamefont{Wang et~al.}(2018)\citenamefont{Wang, Ganose, Niu, and
  Scanlon}}]{WangZ18}
\bibinfo{author}{\bibfnamefont{Z.}~\bibnamefont{Wang}},
  \bibinfo{author}{\bibfnamefont{A.~M.} \bibnamefont{Ganose}},
  \bibinfo{author}{\bibfnamefont{C.}~\bibnamefont{Niu}}, \bibnamefont{and}
  \bibinfo{author}{\bibfnamefont{D.~O.} \bibnamefont{Scanlon}},
  \bibinfo{journal}{J. Mater. Chem. A} \textbf{\bibinfo{volume}{6}},
  \bibinfo{pages}{5652} (\bibinfo{year}{2018}).

\bibitem[{\citenamefont{Liu et~al.}(2018)\citenamefont{Liu, Huhn, Du,
  {Vazquez-Mayagoitia}, Dirkes, You, Kanai, Mitzi, and Blum}}]{LiuC18}
\bibinfo{author}{\bibfnamefont{C.}~\bibnamefont{Liu}},
  \bibinfo{author}{\bibfnamefont{W.}~\bibnamefont{Huhn}},
  \bibinfo{author}{\bibfnamefont{K.-Z.} \bibnamefont{Du}},
  \bibinfo{author}{\bibfnamefont{A.}~\bibnamefont{{Vazquez-Mayagoitia}}},
  \bibinfo{author}{\bibfnamefont{D.}~\bibnamefont{Dirkes}},
  \bibinfo{author}{\bibfnamefont{W.}~\bibnamefont{You}},
  \bibinfo{author}{\bibfnamefont{Y.}~\bibnamefont{Kanai}},
  \bibinfo{author}{\bibfnamefont{D.~B.} \bibnamefont{Mitzi}}, \bibnamefont{and}
  \bibinfo{author}{\bibfnamefont{V.}~\bibnamefont{Blum}},
  \bibinfo{journal}{Phys. Rev. Lett.} \textbf{\bibinfo{volume}{121}},
  \bibinfo{pages}{146401} (\bibinfo{year}{2018}).

\bibitem[{\citenamefont{Ran et~al.}(2019)\citenamefont{Ran, Gao, Li, Xi, Li,
  Dai, Yang, Gao, Dong, Jiao et~al.}}]{Ran19}
\bibinfo{author}{\bibfnamefont{C.}~\bibnamefont{Ran}},
  \bibinfo{author}{\bibfnamefont{W.}~\bibnamefont{Gao}},
  \bibinfo{author}{\bibfnamefont{J.}~\bibnamefont{Li}},
  \bibinfo{author}{\bibfnamefont{J.}~\bibnamefont{Xi}},
  \bibinfo{author}{\bibfnamefont{L.}~\bibnamefont{Li}},
  \bibinfo{author}{\bibfnamefont{J.}~\bibnamefont{Dai}},
  \bibinfo{author}{\bibfnamefont{Y.}~\bibnamefont{Yang}},
  \bibinfo{author}{\bibfnamefont{X.}~\bibnamefont{Gao}},
  \bibinfo{author}{\bibfnamefont{H.}~\bibnamefont{Dong}},
  \bibinfo{author}{\bibfnamefont{B.}~\bibnamefont{Jiao}}, \bibnamefont{et~al.},
  \bibinfo{journal}{Joule} \textbf{\bibinfo{volume}{3}}, \bibinfo{pages}{3072}
  (\bibinfo{year}{2019}).

\bibitem[{\citenamefont{Matteocci et~al.}(2016)\citenamefont{Matteocci, Cin\`a,
  Lamanna, Cacovich, Divitini, Midgley, Ducati, and di~Carlo}}]{Matteocci16}
\bibinfo{author}{\bibfnamefont{F.}~\bibnamefont{Matteocci}},
  \bibinfo{author}{\bibfnamefont{L.}~\bibnamefont{Cin\`a}},
  \bibinfo{author}{\bibfnamefont{E.}~\bibnamefont{Lamanna}},
  \bibinfo{author}{\bibfnamefont{S.}~\bibnamefont{Cacovich}},
  \bibinfo{author}{\bibfnamefont{G.}~\bibnamefont{Divitini}},
  \bibinfo{author}{\bibfnamefont{P.~A.} \bibnamefont{Midgley}},
  \bibinfo{author}{\bibfnamefont{C.}~\bibnamefont{Ducati}}, \bibnamefont{and}
  \bibinfo{author}{\bibfnamefont{A.}~\bibnamefont{di~Carlo}},
  \bibinfo{journal}{Nano Energy} \textbf{\bibinfo{volume}{30}},
  \bibinfo{pages}{162} (\bibinfo{year}{2016}).

\bibitem[{\citenamefont{Cheacharoen
  et~al.}(2018{\natexlab{a}})\citenamefont{Cheacharoen, Rolston, Harwood, Bush,
  Dauskardt, and McGehee}}]{Cheacharoen18a}
\bibinfo{author}{\bibfnamefont{R.}~\bibnamefont{Cheacharoen}},
  \bibinfo{author}{\bibfnamefont{N.}~\bibnamefont{Rolston}},
  \bibinfo{author}{\bibfnamefont{D.}~\bibnamefont{Harwood}},
  \bibinfo{author}{\bibfnamefont{K.~A.} \bibnamefont{Bush}},
  \bibinfo{author}{\bibfnamefont{R.~H.} \bibnamefont{Dauskardt}},
  \bibnamefont{and} \bibinfo{author}{\bibfnamefont{M.~D.}
  \bibnamefont{McGehee}}, \bibinfo{journal}{Energy Environ. Sci.}
  \textbf{\bibinfo{volume}{11}}, \bibinfo{pages}{144}
  (\bibinfo{year}{2018}{\natexlab{a}}).

\bibitem[{\citenamefont{Cheacharoen
  et~al.}(2018{\natexlab{b}})\citenamefont{Cheacharoen, Boyd, Burkhard,
  Leijtens, Raiford, Bush, Bent, and McGehee}}]{Cheacharoen18b}
\bibinfo{author}{\bibfnamefont{R.}~\bibnamefont{Cheacharoen}},
  \bibinfo{author}{\bibfnamefont{C.~C.} \bibnamefont{Boyd}},
  \bibinfo{author}{\bibfnamefont{G.~F.} \bibnamefont{Burkhard}},
  \bibinfo{author}{\bibfnamefont{T.}~\bibnamefont{Leijtens}},
  \bibinfo{author}{\bibfnamefont{J.~A.} \bibnamefont{Raiford}},
  \bibinfo{author}{\bibfnamefont{K.~A.} \bibnamefont{Bush}},
  \bibinfo{author}{\bibfnamefont{S.~F.} \bibnamefont{Bent}}, \bibnamefont{and}
  \bibinfo{author}{\bibfnamefont{M.~D.} \bibnamefont{McGehee}},
  \bibinfo{journal}{Sustain. Energy Fuels} \textbf{\bibinfo{volume}{2}},
  \bibinfo{pages}{2398} (\bibinfo{year}{2018}{\natexlab{b}}).

\bibitem[{\citenamefont{Seidu et~al.}(2019)\citenamefont{Seidu, Himanen, Li,
  and Rinke}}]{Seidu19}
\bibinfo{author}{\bibfnamefont{A.}~\bibnamefont{Seidu}},
  \bibinfo{author}{\bibfnamefont{L.}~\bibnamefont{Himanen}},
  \bibinfo{author}{\bibfnamefont{J.}~\bibnamefont{Li}}, \bibnamefont{and}
  \bibinfo{author}{\bibfnamefont{P.}~\bibnamefont{Rinke}},
  \bibinfo{journal}{New J. Phys.} \textbf{\bibinfo{volume}{21}},
  \bibinfo{pages}{083018} (\bibinfo{year}{2019}).

\bibitem[{\citenamefont{Noh et~al.}(2013)\citenamefont{Noh, Im, Heo, Mandal,
  and Seok}}]{noh}
\bibinfo{author}{\bibfnamefont{J.~H.} \bibnamefont{Noh}},
  \bibinfo{author}{\bibfnamefont{S.~H.} \bibnamefont{Im}},
  \bibinfo{author}{\bibfnamefont{J.~H.} \bibnamefont{Heo}},
  \bibinfo{author}{\bibfnamefont{T.~N.} \bibnamefont{Mandal}},
  \bibnamefont{and} \bibinfo{author}{\bibfnamefont{S.~I.} \bibnamefont{Seok}},
  \bibinfo{journal}{Nano Lett.} \textbf{\bibinfo{volume}{13}},
  \bibinfo{pages}{1764} (\bibinfo{year}{2013}).

\bibitem[{\citenamefont{Yi et~al.}(2016)\citenamefont{Yi, Luo, Meloni, Boziki,
  Ashari-Astani, Gr\"atzel, Zakeeruddin, R\"othlisberger, and
  Gr\"atzel}}]{Yi16}
\bibinfo{author}{\bibfnamefont{C.}~\bibnamefont{Yi}},
  \bibinfo{author}{\bibfnamefont{J.}~\bibnamefont{Luo}},
  \bibinfo{author}{\bibfnamefont{S.}~\bibnamefont{Meloni}},
  \bibinfo{author}{\bibfnamefont{A.}~\bibnamefont{Boziki}},
  \bibinfo{author}{\bibfnamefont{N.}~\bibnamefont{Ashari-Astani}},
  \bibinfo{author}{\bibfnamefont{C.}~\bibnamefont{Gr\"atzel}},
  \bibinfo{author}{\bibfnamefont{S.~M.} \bibnamefont{Zakeeruddin}},
  \bibinfo{author}{\bibfnamefont{U.}~\bibnamefont{R\"othlisberger}},
  \bibnamefont{and}
  \bibinfo{author}{\bibfnamefont{M.}~\bibnamefont{Gr\"atzel}},
  \bibinfo{journal}{Energy Environ. Sci.} \textbf{\bibinfo{volume}{9}},
  \bibinfo{pages}{656} (\bibinfo{year}{2016}).

\bibitem[{\citenamefont{Zhou et~al.}(2016)\citenamefont{Zhou, Zhou, Chen, Zong,
  Huang, Pang, and Padture}}]{ZhouY16}
\bibinfo{author}{\bibfnamefont{Y.}~\bibnamefont{Zhou}},
  \bibinfo{author}{\bibfnamefont{Z.}~\bibnamefont{Zhou}},
  \bibinfo{author}{\bibfnamefont{M.}~\bibnamefont{Chen}},
  \bibinfo{author}{\bibfnamefont{Y.}~\bibnamefont{Zong}},
  \bibinfo{author}{\bibfnamefont{J.}~\bibnamefont{Huang}},
  \bibinfo{author}{\bibfnamefont{S.}~\bibnamefont{Pang}}, \bibnamefont{and}
  \bibinfo{author}{\bibfnamefont{N.~P.} \bibnamefont{Padture}},
  \bibinfo{journal}{J. Mater. Chem. A} \textbf{\bibinfo{volume}{4}},
  \bibinfo{pages}{17623} (\bibinfo{year}{2016}).

\bibitem[{\citenamefont{Tan et~al.}(2017)\citenamefont{Tan, Jain, Voznyy, Lan,
  de~Arquer, Fan, Quintero-Bermudez, Yuan, Zhang, Zhao et~al.}}]{Tan17}
\bibinfo{author}{\bibfnamefont{H.}~\bibnamefont{Tan}},
  \bibinfo{author}{\bibfnamefont{A.}~\bibnamefont{Jain}},
  \bibinfo{author}{\bibfnamefont{O.}~\bibnamefont{Voznyy}},
  \bibinfo{author}{\bibfnamefont{X.}~\bibnamefont{Lan}},
  \bibinfo{author}{\bibfnamefont{F.~P.~G.} \bibnamefont{de~Arquer}},
  \bibinfo{author}{\bibfnamefont{J.~Z.} \bibnamefont{Fan}},
  \bibinfo{author}{\bibfnamefont{R.}~\bibnamefont{Quintero-Bermudez}},
  \bibinfo{author}{\bibfnamefont{M.}~\bibnamefont{Yuan}},
  \bibinfo{author}{\bibfnamefont{B.}~\bibnamefont{Zhang}},
  \bibinfo{author}{\bibfnamefont{Y.}~\bibnamefont{Zhao}}, \bibnamefont{et~al.},
  \bibinfo{journal}{Science} \textbf{\bibinfo{volume}{355}},
  \bibinfo{pages}{722} (\bibinfo{year}{2017}).

\bibitem[{\citenamefont{Gao et~al.}(2018)\citenamefont{Gao, Ran, Li, Dong,
  Zhang, Lan, Hou, and Wu}}]{Gao18}
\bibinfo{author}{\bibfnamefont{W.}~\bibnamefont{Gao}},
  \bibinfo{author}{\bibfnamefont{C.}~\bibnamefont{Ran}},
  \bibinfo{author}{\bibfnamefont{J.}~\bibnamefont{Li}},
  \bibinfo{author}{\bibfnamefont{H.}~\bibnamefont{Dong}},
  \bibinfo{author}{\bibfnamefont{L.}~\bibnamefont{Zhang}},
  \bibinfo{author}{\bibfnamefont{X.}~\bibnamefont{Lan}},
  \bibinfo{author}{\bibfnamefont{X.}~\bibnamefont{Hou}}, \bibnamefont{and}
  \bibinfo{author}{\bibfnamefont{Z.}~\bibnamefont{Wu}}, \bibinfo{journal}{J.
  Phys. Chem. Lett.} \textbf{\bibinfo{volume}{9}}, \bibinfo{pages}{6999}
  (\bibinfo{year}{2018}).

\bibitem[{\citenamefont{Correa-Baena et~al.}(2017)\citenamefont{Correa-Baena,
  Abate, Saliba, Tress, Jacobsson, Gr{\"a}tzel, and Hagfeldt}}]{Baena17}
\bibinfo{author}{\bibfnamefont{J.-P.} \bibnamefont{Correa-Baena}},
  \bibinfo{author}{\bibfnamefont{A.}~\bibnamefont{Abate}},
  \bibinfo{author}{\bibfnamefont{M.}~\bibnamefont{Saliba}},
  \bibinfo{author}{\bibfnamefont{W.}~\bibnamefont{Tress}},
  \bibinfo{author}{\bibfnamefont{T.~J.} \bibnamefont{Jacobsson}},
  \bibinfo{author}{\bibfnamefont{M.}~\bibnamefont{Gr{\"a}tzel}},
  \bibnamefont{and} \bibinfo{author}{\bibfnamefont{A.}~\bibnamefont{Hagfeldt}},
  \bibinfo{journal}{Energy Environ. Sci.} \textbf{\bibinfo{volume}{10}},
  \bibinfo{pages}{710} (\bibinfo{year}{2017}).

\bibitem[{\citenamefont{Ganose et~al.}(2017)\citenamefont{Ganose, Savory, and
  Scanlon}}]{Ganose17}
\bibinfo{author}{\bibfnamefont{A.~M.} \bibnamefont{Ganose}},
  \bibinfo{author}{\bibfnamefont{C.~N.} \bibnamefont{Savory}},
  \bibnamefont{and} \bibinfo{author}{\bibfnamefont{D.~O.}
  \bibnamefont{Scanlon}}, \bibinfo{journal}{Chem. Commun.}
  \textbf{\bibinfo{volume}{53}}, \bibinfo{pages}{20} (\bibinfo{year}{2017}).

\bibitem[{\citenamefont{Frolova et~al.}(2017)\citenamefont{Frolova, Anokhin,
  Piryazev, Luchkin, Dremova, Stevenson, and Troshin}}]{Forolova17}
\bibinfo{author}{\bibfnamefont{L.~A.} \bibnamefont{Frolova}},
  \bibinfo{author}{\bibfnamefont{D.~V.} \bibnamefont{Anokhin}},
  \bibinfo{author}{\bibfnamefont{A.~A.} \bibnamefont{Piryazev}},
  \bibinfo{author}{\bibfnamefont{S.~Y.} \bibnamefont{Luchkin}},
  \bibinfo{author}{\bibfnamefont{N.~N.} \bibnamefont{Dremova}},
  \bibinfo{author}{\bibfnamefont{K.~J.} \bibnamefont{Stevenson}},
  \bibnamefont{and} \bibinfo{author}{\bibfnamefont{P.~A.}
  \bibnamefont{Troshin}}, \bibinfo{journal}{J. Phys. Chem. Lett.}
  \textbf{\bibinfo{volume}{8}}, \bibinfo{pages}{67} (\bibinfo{year}{2017}).

\bibitem[{\citenamefont{Wang et~al.}(2019)\citenamefont{Wang, Dar, Ono, Zhang,
  Kan, Li, Zhang, Wang, Yang, Gao et~al.}}]{WangY19}
\bibinfo{author}{\bibfnamefont{Y.}~\bibnamefont{Wang}},
  \bibinfo{author}{\bibfnamefont{M.~I.} \bibnamefont{Dar}},
  \bibinfo{author}{\bibfnamefont{L.~K.} \bibnamefont{Ono}},
  \bibinfo{author}{\bibfnamefont{T.}~\bibnamefont{Zhang}},
  \bibinfo{author}{\bibfnamefont{M.}~\bibnamefont{Kan}},
  \bibinfo{author}{\bibfnamefont{Y.}~\bibnamefont{Li}},
  \bibinfo{author}{\bibfnamefont{L.}~\bibnamefont{Zhang}},
  \bibinfo{author}{\bibfnamefont{X.}~\bibnamefont{Wang}},
  \bibinfo{author}{\bibfnamefont{Y.}~\bibnamefont{Yang}},
  \bibinfo{author}{\bibfnamefont{X.}~\bibnamefont{Gao}}, \bibnamefont{et~al.},
  \bibinfo{journal}{Science} \textbf{\bibinfo{volume}{365}},
  \bibinfo{pages}{591} (\bibinfo{year}{2019}).

\bibitem[{\citenamefont{Haruyama et~al.}(2014)\citenamefont{Haruyama, Sodeyama,
  Han, and Tateyama}}]{Haruyama14}
\bibinfo{author}{\bibfnamefont{J.}~\bibnamefont{Haruyama}},
  \bibinfo{author}{\bibfnamefont{K.}~\bibnamefont{Sodeyama}},
  \bibinfo{author}{\bibfnamefont{L.}~\bibnamefont{Han}}, \bibnamefont{and}
  \bibinfo{author}{\bibfnamefont{Y.}~\bibnamefont{Tateyama}},
  \bibinfo{journal}{J. Phys. Chem. Lett.} \textbf{\bibinfo{volume}{5}},
  \bibinfo{pages}{2903} (\bibinfo{year}{2014}).

\bibitem[{\citenamefont{Haruyama et~al.}(2016)\citenamefont{Haruyama, Sodeyama,
  Han, and Tateyama}}]{Haruyama16}
\bibinfo{author}{\bibfnamefont{J.}~\bibnamefont{Haruyama}},
  \bibinfo{author}{\bibfnamefont{K.}~\bibnamefont{Sodeyama}},
  \bibinfo{author}{\bibfnamefont{L.}~\bibnamefont{Han}}, \bibnamefont{and}
  \bibinfo{author}{\bibfnamefont{Y.}~\bibnamefont{Tateyama}},
  \bibinfo{journal}{Acc. Chem. Res.} \textbf{\bibinfo{volume}{49}},
  \bibinfo{pages}{554} (\bibinfo{year}{2016}).

\bibitem[{\citenamefont{Gr{\aa}n{\"a}s
  et~al.}(2016)\citenamefont{Gr{\aa}n{\"a}s, Vinichenko, and
  Kaxiras}}]{Oscar16}
\bibinfo{author}{\bibfnamefont{O.}~\bibnamefont{Gr{\aa}n{\"a}s}},
  \bibinfo{author}{\bibfnamefont{D.}~\bibnamefont{Vinichenko}},
  \bibnamefont{and} \bibinfo{author}{\bibfnamefont{E.}~\bibnamefont{Kaxiras}},
  \bibinfo{journal}{Sci. Rep.} \textbf{\bibinfo{volume}{6}},
  \bibinfo{pages}{36108} (\bibinfo{year}{2016}).

\bibitem[{\citenamefont{Akbari et~al.}(2017)\citenamefont{Akbari, Hashemi,
  Mosconi, Angelis, and Hakala}}]{Akbari17}
\bibinfo{author}{\bibfnamefont{A.}~\bibnamefont{Akbari}},
  \bibinfo{author}{\bibfnamefont{J.}~\bibnamefont{Hashemi}},
  \bibinfo{author}{\bibfnamefont{E.}~\bibnamefont{Mosconi}},
  \bibinfo{author}{\bibfnamefont{F.~D.} \bibnamefont{Angelis}},
  \bibnamefont{and} \bibinfo{author}{\bibfnamefont{M.}~\bibnamefont{Hakala}},
  \bibinfo{journal}{J. Mater. Chem. A} \textbf{\bibinfo{volume}{5}},
  \bibinfo{pages}{2339} (\bibinfo{year}{2017}).

\bibitem[{\citenamefont{Jiang et~al.}(2019)\citenamefont{Jiang, Zhao, Zhang,
  Yang, Chen, Chu, Ye, Li, Yin, and You}}]{Jiang19}
\bibinfo{author}{\bibfnamefont{Q.}~\bibnamefont{Jiang}},
  \bibinfo{author}{\bibfnamefont{Y.}~\bibnamefont{Zhao}},
  \bibinfo{author}{\bibfnamefont{X.}~\bibnamefont{Zhang}},
  \bibinfo{author}{\bibfnamefont{X.}~\bibnamefont{Yang}},
  \bibinfo{author}{\bibfnamefont{Y.}~\bibnamefont{Chen}},
  \bibinfo{author}{\bibfnamefont{Z.}~\bibnamefont{Chu}},
  \bibinfo{author}{\bibfnamefont{Q.}~\bibnamefont{Ye}},
  \bibinfo{author}{\bibfnamefont{X.}~\bibnamefont{Li}},
  \bibinfo{author}{\bibfnamefont{Z.}~\bibnamefont{Yin}}, \bibnamefont{and}
  \bibinfo{author}{\bibfnamefont{J.}~\bibnamefont{You}}, \bibinfo{journal}{Nat.
  Photon.} \textbf{\bibinfo{volume}{13}}, \bibinfo{pages}{460}
  (\bibinfo{year}{2019}).

\bibitem[{\citenamefont{Chen et~al.}(2018)\citenamefont{Chen, Bai, Wang, Lyu,
  Yun, and Wang}}]{Chen18}
\bibinfo{author}{\bibfnamefont{P.}~\bibnamefont{Chen}},
  \bibinfo{author}{\bibfnamefont{Y.}~\bibnamefont{Bai}},
  \bibinfo{author}{\bibfnamefont{S.}~\bibnamefont{Wang}},
  \bibinfo{author}{\bibfnamefont{M.}~\bibnamefont{Lyu}},
  \bibinfo{author}{\bibfnamefont{J.-H.} \bibnamefont{Yun}}, \bibnamefont{and}
  \bibinfo{author}{\bibfnamefont{L.}~\bibnamefont{Wang}},
  \bibinfo{journal}{Adv. Funct. Mater.} \textbf{\bibinfo{volume}{28}},
  \bibinfo{pages}{1706923} (\bibinfo{year}{2018}).

\bibitem[{\citenamefont{Cho et~al.}(2018)\citenamefont{Cho, Soufiani, Yun, Kim,
  Lee, Seidel, Deng, Green, Huang, and Ho-Baillie}}]{Cho18}
\bibinfo{author}{\bibfnamefont{H.}~\bibnamefont{Cho}},
  \bibinfo{author}{\bibfnamefont{A.~M.} \bibnamefont{Soufiani}},
  \bibinfo{author}{\bibfnamefont{J.~S.} \bibnamefont{Yun}},
  \bibinfo{author}{\bibfnamefont{J.}~\bibnamefont{Kim}},
  \bibinfo{author}{\bibfnamefont{D.~S.} \bibnamefont{Lee}},
  \bibinfo{author}{\bibfnamefont{J.}~\bibnamefont{Seidel}},
  \bibinfo{author}{\bibfnamefont{X.}~\bibnamefont{Deng}},
  \bibinfo{author}{\bibfnamefont{M.~A.} \bibnamefont{Green}},
  \bibinfo{author}{\bibfnamefont{S.}~\bibnamefont{Huang}}, \bibnamefont{and}
  \bibinfo{author}{\bibfnamefont{A.~W.~Y.} \bibnamefont{Ho-Baillie}},
  \bibinfo{journal}{Adv. Energy. Mater.} \textbf{\bibinfo{volume}{8}},
  \bibinfo{pages}{1703392} (\bibinfo{year}{2018}).

\bibitem[{\citenamefont{Saliba et~al.}(2016{\natexlab{a}})\citenamefont{Saliba,
  Matsui, Seo, Domanski, Correa-Baena, Nazeeruddin, Zakeeruddin, Tress, Abate,
  Hagfeldt et~al.}}]{Saliba16a}
\bibinfo{author}{\bibfnamefont{M.}~\bibnamefont{Saliba}},
  \bibinfo{author}{\bibfnamefont{T.}~\bibnamefont{Matsui}},
  \bibinfo{author}{\bibfnamefont{J.-Y.} \bibnamefont{Seo}},
  \bibinfo{author}{\bibfnamefont{K.}~\bibnamefont{Domanski}},
  \bibinfo{author}{\bibfnamefont{J.-P.} \bibnamefont{Correa-Baena}},
  \bibinfo{author}{\bibfnamefont{M.~K.} \bibnamefont{Nazeeruddin}},
  \bibinfo{author}{\bibfnamefont{S.~M.} \bibnamefont{Zakeeruddin}},
  \bibinfo{author}{\bibfnamefont{W.}~\bibnamefont{Tress}},
  \bibinfo{author}{\bibfnamefont{A.}~\bibnamefont{Abate}},
  \bibinfo{author}{\bibfnamefont{A.}~\bibnamefont{Hagfeldt}},
  \bibnamefont{et~al.}, \bibinfo{journal}{Energy Environ. Sci.}
  \textbf{\bibinfo{volume}{9}}, \bibinfo{pages}{1989}
  (\bibinfo{year}{2016}{\natexlab{a}}).

\bibitem[{\citenamefont{Saliba et~al.}(2016{\natexlab{b}})\citenamefont{Saliba,
  Orlandi, Matsui, Aghazada, Cavazzini, Correa-Baena, Gao, Scopelliti, Mosconi,
  Dahmen et~al.}}]{Saliba16b}
\bibinfo{author}{\bibfnamefont{M.}~\bibnamefont{Saliba}},
  \bibinfo{author}{\bibfnamefont{S.}~\bibnamefont{Orlandi}},
  \bibinfo{author}{\bibfnamefont{T.}~\bibnamefont{Matsui}},
  \bibinfo{author}{\bibfnamefont{S.}~\bibnamefont{Aghazada}},
  \bibinfo{author}{\bibfnamefont{M.}~\bibnamefont{Cavazzini}},
  \bibinfo{author}{\bibfnamefont{J.-P.} \bibnamefont{Correa-Baena}},
  \bibinfo{author}{\bibfnamefont{P.}~\bibnamefont{Gao}},
  \bibinfo{author}{\bibfnamefont{R.}~\bibnamefont{Scopelliti}},
  \bibinfo{author}{\bibfnamefont{E.}~\bibnamefont{Mosconi}},
  \bibinfo{author}{\bibfnamefont{K.-H.} \bibnamefont{Dahmen}},
  \bibnamefont{et~al.}, \bibinfo{journal}{Nat. Energy}
  \textbf{\bibinfo{volume}{1}}, \bibinfo{pages}{15017}
  (\bibinfo{year}{2016}{\natexlab{b}}).

\bibitem[{\citenamefont{Saliba et~al.}(2016{\natexlab{c}})\citenamefont{Saliba,
  Matsui, Domanski, Seo, Ummadisingu, Zakeeruddin, Correa-Baena, Tress, Abate,
  Hagfeldt et~al.}}]{Saliba16c}
\bibinfo{author}{\bibfnamefont{M.}~\bibnamefont{Saliba}},
  \bibinfo{author}{\bibfnamefont{T.}~\bibnamefont{Matsui}},
  \bibinfo{author}{\bibfnamefont{K.}~\bibnamefont{Domanski}},
  \bibinfo{author}{\bibfnamefont{J.-Y.} \bibnamefont{Seo}},
  \bibinfo{author}{\bibfnamefont{A.}~\bibnamefont{Ummadisingu}},
  \bibinfo{author}{\bibfnamefont{S.~M.} \bibnamefont{Zakeeruddin}},
  \bibinfo{author}{\bibfnamefont{J.-P.} \bibnamefont{Correa-Baena}},
  \bibinfo{author}{\bibfnamefont{W.}~\bibnamefont{Tress}},
  \bibinfo{author}{\bibfnamefont{A.}~\bibnamefont{Abate}},
  \bibinfo{author}{\bibfnamefont{A.}~\bibnamefont{Hagfeldt}},
  \bibnamefont{et~al.}, \bibinfo{journal}{Science}
  \textbf{\bibinfo{volume}{354}}, \bibinfo{pages}{206}
  (\bibinfo{year}{2016}{\natexlab{c}}).

\bibitem[{\citenamefont{Li et~al.}(2017{\natexlab{a}})\citenamefont{Li, Zhang,
  Huang, Shen, Cheng, and Huang}}]{Li17}
\bibinfo{author}{\bibfnamefont{Y.}~\bibnamefont{Li}},
  \bibinfo{author}{\bibfnamefont{C.}~\bibnamefont{Zhang}},
  \bibinfo{author}{\bibfnamefont{D.}~\bibnamefont{Huang}},
  \bibinfo{author}{\bibfnamefont{Q.}~\bibnamefont{Shen}},
  \bibinfo{author}{\bibfnamefont{Y.}~\bibnamefont{Cheng}}, \bibnamefont{and}
  \bibinfo{author}{\bibfnamefont{W.}~\bibnamefont{Huang}},
  \bibinfo{journal}{Appl. Phys. Lett.} \textbf{\bibinfo{volume}{111}},
  \bibinfo{pages}{162106} (\bibinfo{year}{2017}{\natexlab{a}}).

\bibitem[{\citenamefont{Huang et~al.}(2018)\citenamefont{Huang, Yin, and
  He}}]{Huang18}
\bibinfo{author}{\bibfnamefont{Y.}~\bibnamefont{Huang}},
  \bibinfo{author}{\bibfnamefont{W.-J.} \bibnamefont{Yin}}, \bibnamefont{and}
  \bibinfo{author}{\bibfnamefont{Y.}~\bibnamefont{He}}, \bibinfo{journal}{J.
  Phys. Chem. C} pp. \bibinfo{pages}{1345--1350} (\bibinfo{year}{2018}).

\bibitem[{\citenamefont{Sutton et~al.}(2018)\citenamefont{Sutton, R.Filip,
  Haghighirad, Sakai, Wenger, Giustino, and Snaith}}]{Sutton18}
\bibinfo{author}{\bibfnamefont{R.~J.} \bibnamefont{Sutton}},
  \bibinfo{author}{\bibfnamefont{M.}~\bibnamefont{R.Filip}},
  \bibinfo{author}{\bibfnamefont{A.~A.} \bibnamefont{Haghighirad}},
  \bibinfo{author}{\bibfnamefont{N.}~\bibnamefont{Sakai}},
  \bibinfo{author}{\bibfnamefont{B.}~\bibnamefont{Wenger}},
  \bibinfo{author}{\bibfnamefont{F.}~\bibnamefont{Giustino}}, \bibnamefont{and}
  \bibinfo{author}{\bibfnamefont{H.~J.} \bibnamefont{Snaith}},
  \bibinfo{journal}{ACS Energy Lett.} \textbf{\bibinfo{volume}{3}},
  \bibinfo{pages}{1787} (\bibinfo{year}{2018}).

\bibitem[{\citenamefont{Evarestov et~al.}(2019)\citenamefont{Evarestov,
  Sanocrate, Kotomin, and Maier}}]{Evarestov19}
\bibinfo{author}{\bibfnamefont{R.~A.} \bibnamefont{Evarestov}},
  \bibinfo{author}{\bibfnamefont{A.}~\bibnamefont{Sanocrate}},
  \bibinfo{author}{\bibfnamefont{E.~A.} \bibnamefont{Kotomin}},
  \bibnamefont{and} \bibinfo{author}{\bibfnamefont{J.}~\bibnamefont{Maier}},
  \bibinfo{journal}{Phys. Chem. Chem. Phys.} \textbf{\bibinfo{volume}{21}},
  \bibinfo{pages}{7841} (\bibinfo{year}{2019}).

\bibitem[{\citenamefont{Reuter and Scheffler}(2003{\natexlab{a}})}]{karsten}
\bibinfo{author}{\bibfnamefont{K.}~\bibnamefont{Reuter}} \bibnamefont{and}
  \bibinfo{author}{\bibfnamefont{M.}~\bibnamefont{Scheffler}},
  \bibinfo{journal}{Phys. Rev. Lett.} \textbf{\bibinfo{volume}{90}},
  \bibinfo{pages}{0461031} (\bibinfo{year}{2003}{\natexlab{a}}).

\bibitem[{\citenamefont{{Karsten Reuter and Catherine Stampfl and Matthias
  Scheffler}}(2005)}]{reuter}
\bibinfo{author}{\bibnamefont{{Karsten Reuter and Catherine Stampfl and
  Matthias Scheffler}}}, \emph{\bibinfo{title}{Handbook of {Material
  Modeling}}} (\bibinfo{publisher}{{Springer Dordreccht}},
  \bibinfo{year}{2005}), \bibinfo{edition}{s. yip} ed.

\bibitem[{\citenamefont{Reuter and
  Scheffler}(2003{\natexlab{b}})}]{abinitiothermodynamics}
\bibinfo{author}{\bibfnamefont{K.}~\bibnamefont{Reuter}} \bibnamefont{and}
  \bibinfo{author}{\bibfnamefont{M.}~\bibnamefont{Scheffler}},
  \bibinfo{journal}{Phys. Rev. B} \textbf{\bibinfo{volume}{68}},
  \bibinfo{pages}{045407} (\bibinfo{year}{2003}{\natexlab{b}}).

\bibitem[{\citenamefont{Perdew et~al.}(2008)\citenamefont{Perdew, Ruzsinszky,
  Csonka, Vydrov, Scuseria, Constantin, Zhou, and Burke}}]{perdew08}
\bibinfo{author}{\bibfnamefont{J.~P.} \bibnamefont{Perdew}},
  \bibinfo{author}{\bibfnamefont{A.}~\bibnamefont{Ruzsinszky}},
  \bibinfo{author}{\bibfnamefont{G.~I.} \bibnamefont{Csonka}},
  \bibinfo{author}{\bibfnamefont{O.~A.} \bibnamefont{Vydrov}},
  \bibinfo{author}{\bibfnamefont{G.~E.} \bibnamefont{Scuseria}},
  \bibinfo{author}{\bibfnamefont{L.~A.} \bibnamefont{Constantin}},
  \bibinfo{author}{\bibfnamefont{X.}~\bibnamefont{Zhou}}, \bibnamefont{and}
  \bibinfo{author}{\bibfnamefont{K.}~\bibnamefont{Burke}},
  \bibinfo{journal}{Phys. Rev. Lett.} \textbf{\bibinfo{volume}{100}},
  \bibinfo{pages}{136406} (\bibinfo{year}{2008}).

\bibitem[{\citenamefont{Blum et~al.}(2009)\citenamefont{Blum, Gehrke, Hanke,
  Havu, Havu, Ren, Reuter, and Scheffler}}]{Blum09}
\bibinfo{author}{\bibfnamefont{V.}~\bibnamefont{Blum}},
  \bibinfo{author}{\bibfnamefont{R.}~\bibnamefont{Gehrke}},
  \bibinfo{author}{\bibfnamefont{F.}~\bibnamefont{Hanke}},
  \bibinfo{author}{\bibfnamefont{P.}~\bibnamefont{Havu}},
  \bibinfo{author}{\bibfnamefont{V.}~\bibnamefont{Havu}},
  \bibinfo{author}{\bibfnamefont{X.}~\bibnamefont{Ren}},
  \bibinfo{author}{\bibfnamefont{K.}~\bibnamefont{Reuter}}, \bibnamefont{and}
  \bibinfo{author}{\bibfnamefont{M.}~\bibnamefont{Scheffler}},
  \bibinfo{journal}{Comput. Phys. Commun.} \textbf{\bibinfo{volume}{180}},
  \bibinfo{pages}{2175} (\bibinfo{year}{2009}).

\bibitem[{\citenamefont{Havu et~al.}(2009)\citenamefont{Havu, Blum, Havu, and
  Scheffler}}]{havu}
\bibinfo{author}{\bibfnamefont{V.}~\bibnamefont{Havu}},
  \bibinfo{author}{\bibfnamefont{V.}~\bibnamefont{Blum}},
  \bibinfo{author}{\bibfnamefont{P.}~\bibnamefont{Havu}}, \bibnamefont{and}
  \bibinfo{author}{\bibfnamefont{M.}~\bibnamefont{Scheffler}},
  \bibinfo{journal}{J. Comput. Phys.} \textbf{\bibinfo{volume}{228}},
  \bibinfo{pages}{8367} (\bibinfo{year}{2009}).

\bibitem[{\citenamefont{Levchenko et~al.}(2015)\citenamefont{Levchenko, Ren,
  Wieferink, Johanni, Rinke, Blum, and Scheffler}}]{Levchenko/etal:2015}
\bibinfo{author}{\bibfnamefont{S.~V.} \bibnamefont{Levchenko}},
  \bibinfo{author}{\bibfnamefont{X.}~\bibnamefont{Ren}},
  \bibinfo{author}{\bibfnamefont{J.}~\bibnamefont{Wieferink}},
  \bibinfo{author}{\bibfnamefont{R.}~\bibnamefont{Johanni}},
  \bibinfo{author}{\bibfnamefont{P.}~\bibnamefont{Rinke}},
  \bibinfo{author}{\bibfnamefont{V.}~\bibnamefont{Blum}}, \bibnamefont{and}
  \bibinfo{author}{\bibfnamefont{M.}~\bibnamefont{Scheffler}},
  \bibinfo{journal}{Comput. Phys. Commun.} \textbf{\bibinfo{volume}{192}},
  \bibinfo{pages}{60 } (\bibinfo{year}{2015}).

\bibitem[{\citenamefont{Yang et~al.}(2017)\citenamefont{Yang, Skelton,
  da~Silva, Frost, and Walsh}}]{Yang17}
\bibinfo{author}{\bibfnamefont{R.~X.} \bibnamefont{Yang}},
  \bibinfo{author}{\bibfnamefont{J.~M.} \bibnamefont{Skelton}},
  \bibinfo{author}{\bibfnamefont{E.~L.} \bibnamefont{da~Silva}},
  \bibinfo{author}{\bibfnamefont{J.~M.} \bibnamefont{Frost}}, \bibnamefont{and}
  \bibinfo{author}{\bibfnamefont{A.}~\bibnamefont{Walsh}}, \bibinfo{journal}{J.
  Phys. Chem. Lett.} \textbf{\bibinfo{volume}{8}}, \bibinfo{pages}{4720}
  (\bibinfo{year}{2017}).

\bibitem[{\citenamefont{Bokdam et~al.}(2017)\citenamefont{Bokdam, Lahnsteiner,
  Ramberger, Sch{\"a}fer, and Kresse}}]{Bokdam17}
\bibinfo{author}{\bibfnamefont{M.}~\bibnamefont{Bokdam}},
  \bibinfo{author}{\bibfnamefont{J.}~\bibnamefont{Lahnsteiner}},
  \bibinfo{author}{\bibfnamefont{B.}~\bibnamefont{Ramberger}},
  \bibinfo{author}{\bibfnamefont{T.}~\bibnamefont{Sch{\"a}fer}},
  \bibnamefont{and} \bibinfo{author}{\bibfnamefont{G.}~\bibnamefont{Kresse}},
  \bibinfo{journal}{Phys. Rev. Lett.} \textbf{\bibinfo{volume}{119}},
  \bibinfo{pages}{145501} (\bibinfo{year}{2017}).

\bibitem[{\citenamefont{van Lenthe et~al.}(1993)\citenamefont{van Lenthe,
  Baerends, and Snijders}}]{vanlenthe}
\bibinfo{author}{\bibfnamefont{E.}~\bibnamefont{van Lenthe}},
  \bibinfo{author}{\bibfnamefont{E.~J.} \bibnamefont{Baerends}},
  \bibnamefont{and} \bibinfo{author}{\bibfnamefont{J.~G.}
  \bibnamefont{Snijders}}, \bibinfo{journal}{J. Chem. Phys.}
  \textbf{\bibinfo{volume}{99}} (\bibinfo{year}{1993}).

\bibitem[{\citenamefont{Knuth et~al.}(2015)\citenamefont{Knuth, Carbogno,
  Atalla, Blum, and Scheffler}}]{knuth}
\bibinfo{author}{\bibfnamefont{F.}~\bibnamefont{Knuth}},
  \bibinfo{author}{\bibfnamefont{C.}~\bibnamefont{Carbogno}},
  \bibinfo{author}{\bibfnamefont{V.}~\bibnamefont{Atalla}},
  \bibinfo{author}{\bibfnamefont{V.}~\bibnamefont{Blum}}, \bibnamefont{and}
  \bibinfo{author}{\bibfnamefont{M.}~\bibnamefont{Scheffler}},
  \bibinfo{journal}{Comput. Phys. Commun.} \textbf{\bibinfo{volume}{190}},
  \bibinfo{pages}{33} (\bibinfo{year}{2015}).

\bibitem[{\citenamefont{Neugebauer and Scheffler}(1992)}]{Neugebauer92}
\bibinfo{author}{\bibfnamefont{J.}~\bibnamefont{Neugebauer}} \bibnamefont{and}
  \bibinfo{author}{\bibfnamefont{M.}~\bibnamefont{Scheffler}},
  \bibinfo{journal}{Phys. Rev. B} \textbf{\bibinfo{volume}{46}},
  \bibinfo{pages}{16067} (\bibinfo{year}{1992}).

\bibitem[{\citenamefont{Schulz et~al.}(2019)\citenamefont{Schulz, Cahen, and
  Kahn}}]{Schulz19}
\bibinfo{author}{\bibfnamefont{P.}~\bibnamefont{Schulz}},
  \bibinfo{author}{\bibfnamefont{D.}~\bibnamefont{Cahen}}, \bibnamefont{and}
  \bibinfo{author}{\bibfnamefont{A.}~\bibnamefont{Kahn}},
  \bibinfo{journal}{Chem. Rev.} pp. \bibinfo{pages}{3349--3417}
  (\bibinfo{year}{2019}).

\bibitem[{\citenamefont{Himanen et~al.}(2019)\citenamefont{Himanen, Geurts,
  Foster, and Rinke}}]{Himanen2019}
\bibinfo{author}{\bibfnamefont{L.}~\bibnamefont{Himanen}},
  \bibinfo{author}{\bibfnamefont{A.}~\bibnamefont{Geurts}},
  \bibinfo{author}{\bibfnamefont{A.~S.} \bibnamefont{Foster}},
  \bibnamefont{and} \bibinfo{author}{\bibfnamefont{P.}~\bibnamefont{Rinke}},
  \bibinfo{journal}{Adv. Sci.} \textbf{\bibinfo{volume}{6}},
  \bibinfo{pages}{1900808} (\bibinfo{year}{2019}).

\bibitem[{NOM()}]{NOMAD}
\bibinfo{note}{To avoid the generation of multiple DOIs, submission to NOMAD
  repository will follow inclusion of reviewer's comments.}

\bibitem[{\citenamefont{Yin et~al.}(2014)\citenamefont{Yin, Shi, and
  Yan}}]{Yin14}
\bibinfo{author}{\bibfnamefont{W.-J.} \bibnamefont{Yin}},
  \bibinfo{author}{\bibfnamefont{T.}~\bibnamefont{Shi}}, \bibnamefont{and}
  \bibinfo{author}{\bibfnamefont{Y.}~\bibnamefont{Yan}},
  \bibinfo{journal}{Appl. Phys. Lett.} \textbf{\bibinfo{volume}{104}},
  \bibinfo{pages}{063903} (\bibinfo{year}{2014}).

\bibitem[{\citenamefont{Geng et~al.}(2015)\citenamefont{Geng, Tong, Tang, Yam,
  Zhang, Lau, and Liu}}]{Geng15}
\bibinfo{author}{\bibfnamefont{W.}~\bibnamefont{Geng}},
  \bibinfo{author}{\bibfnamefont{C.-J.} \bibnamefont{Tong}},
  \bibinfo{author}{\bibfnamefont{Z.-K.} \bibnamefont{Tang}},
  \bibinfo{author}{\bibfnamefont{C.}~\bibnamefont{Yam}},
  \bibinfo{author}{\bibfnamefont{Y.-N.} \bibnamefont{Zhang}},
  \bibinfo{author}{\bibfnamefont{W.-M.} \bibnamefont{Lau}}, \bibnamefont{and}
  \bibinfo{author}{\bibfnamefont{L.-M.} \bibnamefont{Liu}},
  \bibinfo{journal}{Journal of Materiomics} \textbf{\bibinfo{volume}{1}},
  \bibinfo{pages}{213} (\bibinfo{year}{2015}).

\bibitem[{\citenamefont{Haruyama et~al.}(2015)\citenamefont{Haruyama, Sodeyama,
  Han, and Tateyama}}]{Haruyama15}
\bibinfo{author}{\bibfnamefont{J.}~\bibnamefont{Haruyama}},
  \bibinfo{author}{\bibfnamefont{K.}~\bibnamefont{Sodeyama}},
  \bibinfo{author}{\bibfnamefont{L.}~\bibnamefont{Han}}, \bibnamefont{and}
  \bibinfo{author}{\bibfnamefont{Y.}~\bibnamefont{Tateyama}},
  \bibinfo{journal}{J. Am. Chem. Soc.} \textbf{\bibinfo{volume}{137}},
  \bibinfo{pages}{10048} (\bibinfo{year}{2015}).

\bibitem[{\citenamefont{Gurvich et~al.}(1989)\citenamefont{Gurvich, Veyts, and
  Alcock}}]{NIST}
\bibinfo{author}{\bibfnamefont{L.~V.} \bibnamefont{Gurvich}},
  \bibinfo{author}{\bibfnamefont{I.~V.} \bibnamefont{Veyts}}, \bibnamefont{and}
  \bibinfo{author}{\bibfnamefont{C.~B.} \bibnamefont{Alcock}},
  \emph{\bibinfo{title}{Thermodynamic Properties of Individual Substances}}
  (\bibinfo{publisher}{Hemisphere Pub. Co. New York}, \bibinfo{year}{1989}).

\bibitem[{\citenamefont{Ingesfield}(1982)}]{Inglesfield82}
\bibinfo{author}{\bibfnamefont{J.~E.} \bibnamefont{Ingesfield}},
  \bibinfo{journal}{Rep. Prog. Phys.} \textbf{\bibinfo{volume}{45}},
  \bibinfo{pages}{223} (\bibinfo{year}{1982}).

\bibitem[{\citenamefont{Speer et~al.}(2009)\citenamefont{Speer, Brinkley, Liu,
  Wei, Miller, and Chiang}}]{Speer09}
\bibinfo{author}{\bibfnamefont{N.~J.} \bibnamefont{Speer}},
  \bibinfo{author}{\bibfnamefont{M.~K.} \bibnamefont{Brinkley}},
  \bibinfo{author}{\bibfnamefont{Y.}~\bibnamefont{Liu}},
  \bibinfo{author}{\bibfnamefont{C.~M.} \bibnamefont{Wei}},
  \bibinfo{author}{\bibfnamefont{T.}~\bibnamefont{Miller}}, \bibnamefont{and}
  \bibinfo{author}{\bibfnamefont{T.~C.} \bibnamefont{Chiang}},
  \bibinfo{journal}{Europhys. Lett.} \textbf{\bibinfo{volume}{88}},
  \bibinfo{pages}{67004} (\bibinfo{year}{2009}).

\bibitem[{\citenamefont{Li et~al.}(2017{\natexlab{b}})\citenamefont{Li, Liu,
  Bai, Zhang, and Prezhdo}}]{Wei17}
\bibinfo{author}{\bibfnamefont{W.}~\bibnamefont{Li}},
  \bibinfo{author}{\bibfnamefont{J.}~\bibnamefont{Liu}},
  \bibinfo{author}{\bibfnamefont{F.-Q.} \bibnamefont{Bai}},
  \bibinfo{author}{\bibfnamefont{H.-X.} \bibnamefont{Zhang}}, \bibnamefont{and}
  \bibinfo{author}{\bibfnamefont{O.~V.} \bibnamefont{Prezhdo}},
  \bibinfo{journal}{ACS Energy Lett.} \textbf{\bibinfo{volume}{2}},
  \bibinfo{pages}{1270} (\bibinfo{year}{2017}{\natexlab{b}}).

\end{thebibliography}

\end{document}


\title
{{\normalsize{Supplemental Material for}} \\ ~ \\ Atomic and electronic structure of cesium lead triiodide surfaces}

\author{Azimatu Seidu}
\affiliation{Department of Applied Physics, Aalto University, FI-00076 AALTO, Finland}
\author{Marc Dvorak}
\affiliation{Department of Applied Physics, Aalto University, FI-00076 AALTO, Finland}
\author{Patrick Rinke}
\affiliation{Department of Applied Physics, Aalto University, FI-00076 AALTO, Finland}
\author{Jingrui Li}
\affiliation{Electronic Materials Research Laboratory, Key Laboratory of the Ministry of Education \& International Center for Dielectric Research, School of Electronic Science and Engineering, Xi'an Jiaotong University, Xi'an 710049, China}

\maketitle

\newpage

\section{Structures of \texorpdfstring{$\text{CsI}$}{}-T reconstructed models}

\begin{figure}[!htp]
\centering
\includegraphics[width=\linewidth]{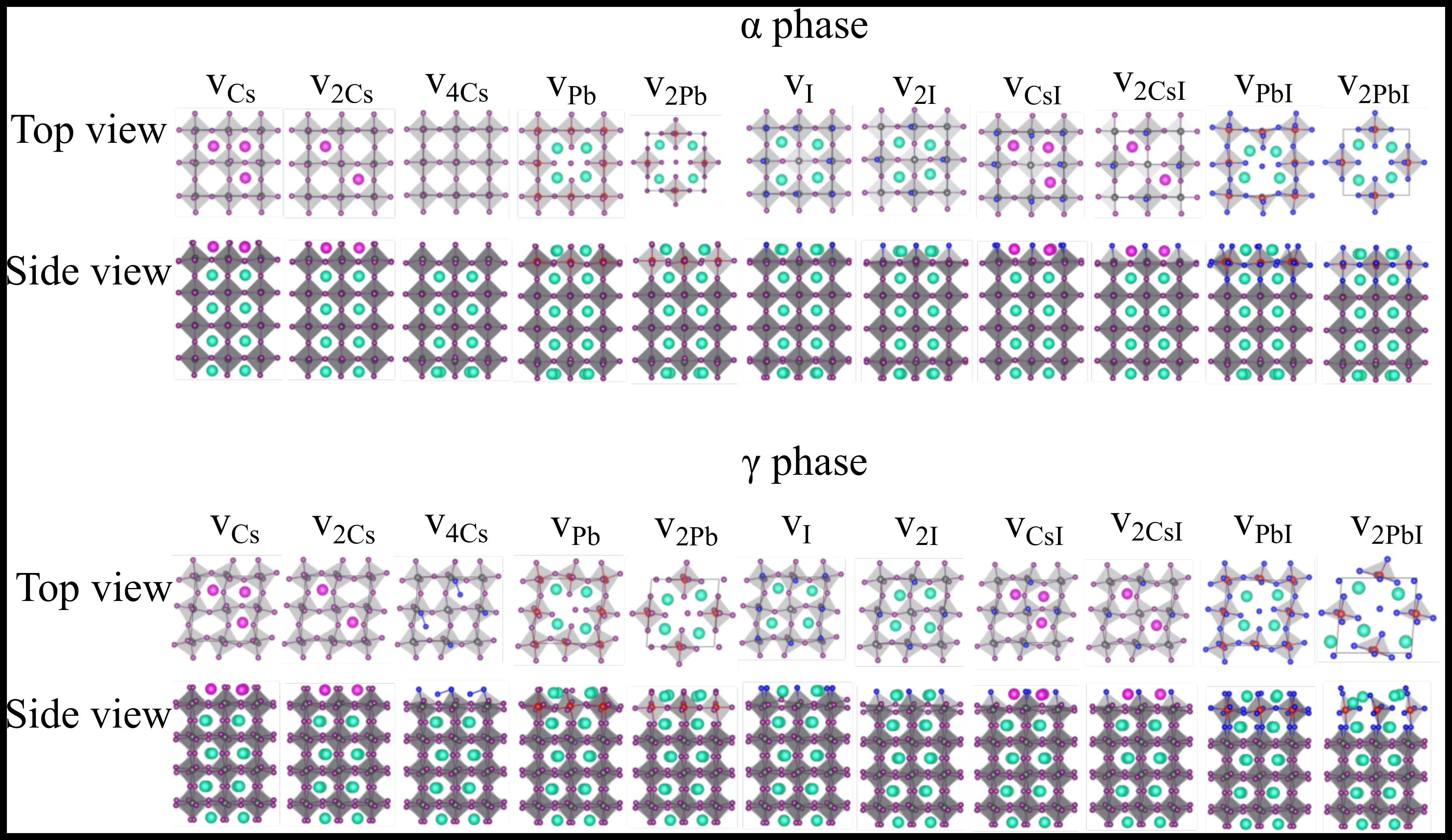}\\\vspace{0.5em}
\includegraphics[width=\linewidth]{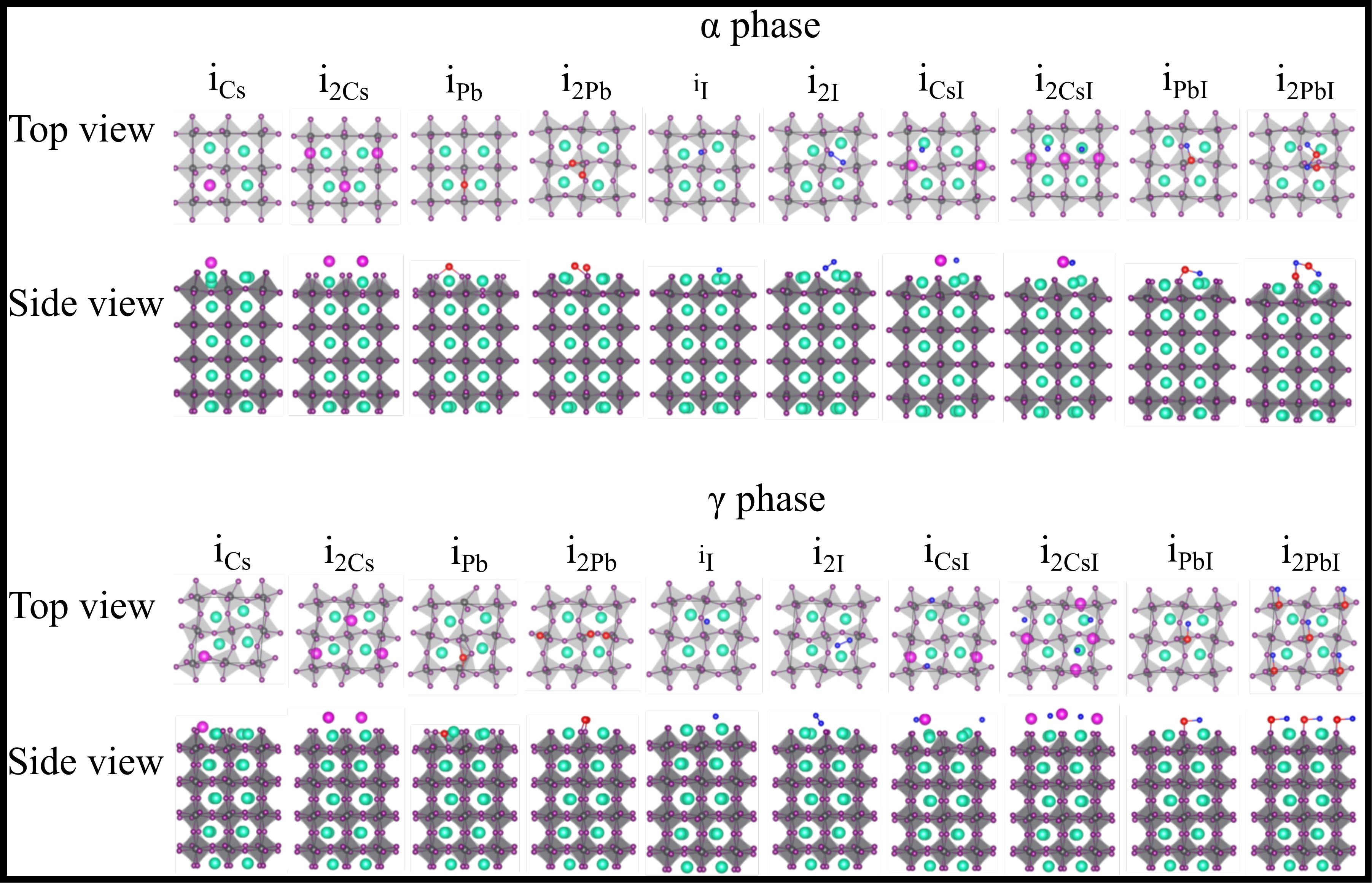}\vspace{-1.0em}
\caption{Structures of reconstructed models of CsI-T surfaces with missing and add atoms as well as complexes in $\upalpha$ and $\upgamma$ phases.}\label{csi}
\end{figure}

Figure~\ref{csi} depicts relaxed structures of all CsI-T models that do not intersect the stable bulk. Surface atoms of interest are highlighted in pink, red and blue for Cs, Pb and I respectively. Surface models with missing and add atoms (complexes) are shown in the top and bottom panels respectively. As alluded to in Sec.~III~A~2, there were slight shifts in atomic positions to accommodate the missing or add atoms (as well as complexes). Additionally, we observe tilting in surface octahedra due to missing or add atoms (complexes) (v$_{\text{PbI}}$ and v$_{2\text{PbI}}$). Notable among the changes in atomic positions are I migrations as discussed in Sec.~III~A~2. The migrated I atoms in i$_{2\text{I}}$ in both phases ($\upalpha$ and $\upgamma$) and $\upgamma$-v$_{4\text{Cs}}$ form I$_2$ molecule in the gaseous phase.
Despite these changes, all the models preserve the CsPbI$_3$ bulk crystal structure. 

\newpage

\section{Charge densities of most relevant reconstruction models in \texorpdfstring{$\text{CsI}$}{}-T surfaces}

\begin{figure*}[!htp]
\includegraphics[width=\linewidth]{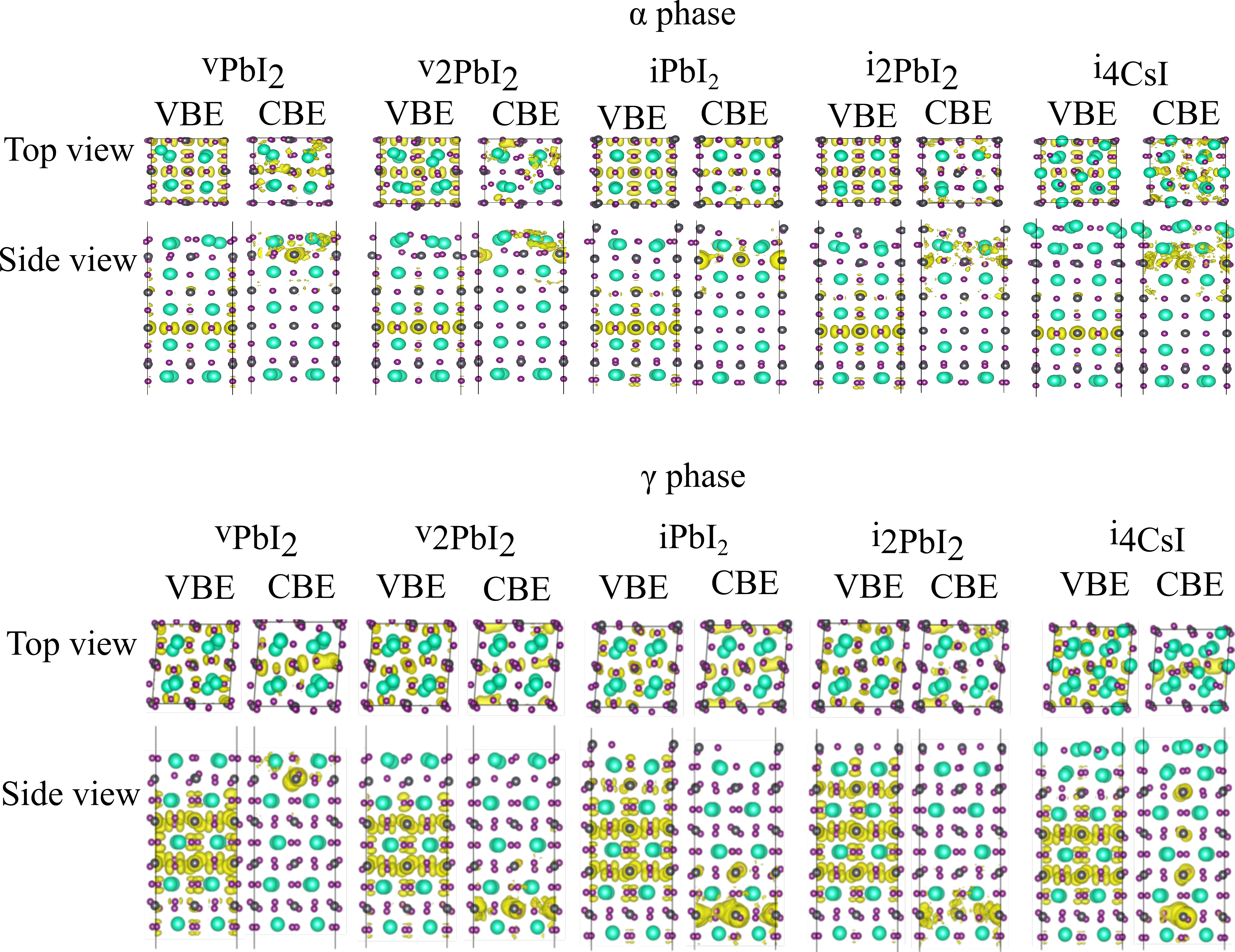}
\caption{Charge distributions of the valence and conduction bands at M of $\upalpha$ (upper panel) and $\upgamma$ (bottom panel).  VBE and CBE denote valence band and conduction band edge respectively.} \label{cube_surf} 
\end{figure*}

The charge distribution of the most relevant CsI-T surface models (v$_{\text{PbI}_2^{}}$, v$_{2\text{PbI}_2^{}}$, i$_{\text{PbI}_2^{}}$, i$_{2\text{PbI}_2^{}}$ and i$_{4\text{CsI}}$) at $k$-points ``M`` (upper panel) and ``$\Gamma$`` (bottom panel) for the $\upalpha$ and the $\upgamma$ phases are depicted in Fig.~\ref{cube_surf}. VBE, CBE, VBM and CBM denote valence band edge, conduction band edges, valence band maximum and conduction band minimum respectively.  

\begin{figure*}[!htp]
\includegraphics[width=\linewidth]{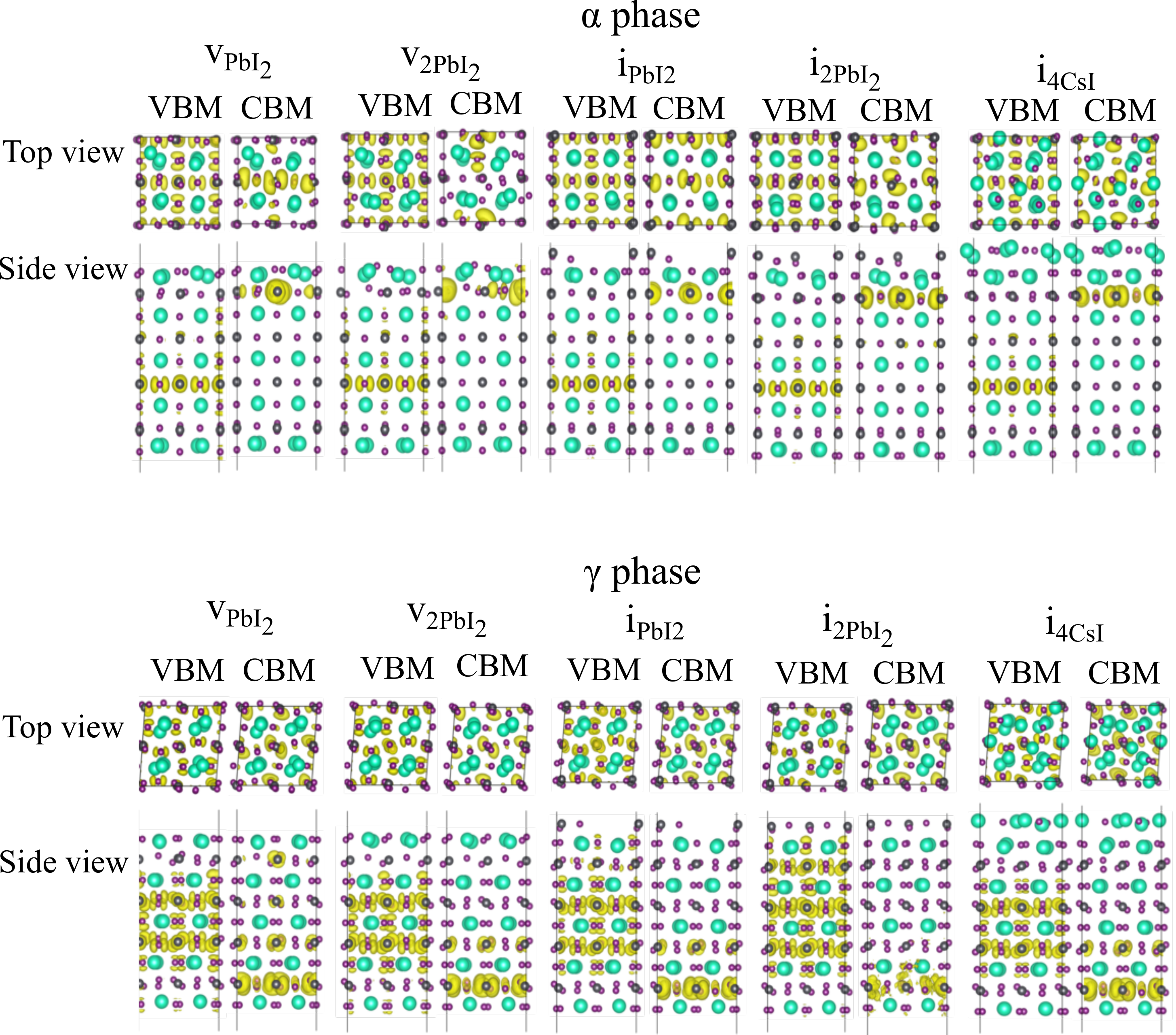}
\caption{Charge distributions of the valence and conduction bands  at $\Gamma$ of $\upalpha$ (upper panel) and $\upgamma$ (bottom panel). VBM and CBM denote valence band maximum and conduction band minimum respectively.} \label{cube_surf_1} 
\end{figure*}

\newpage

\section{\texorpdfstring{$\text{PbI}_2^{}$}{}-T surfaces}

This section highlights the results from PbI$_2$-T surfaces models. Here, we show the surface phase diagrams and crystal structures of all the models we investigated.

\subsection{Surface phase diagrams}\label{SM}

\begin{figure}[!htp]
\centering
\includegraphics[width=\linewidth]{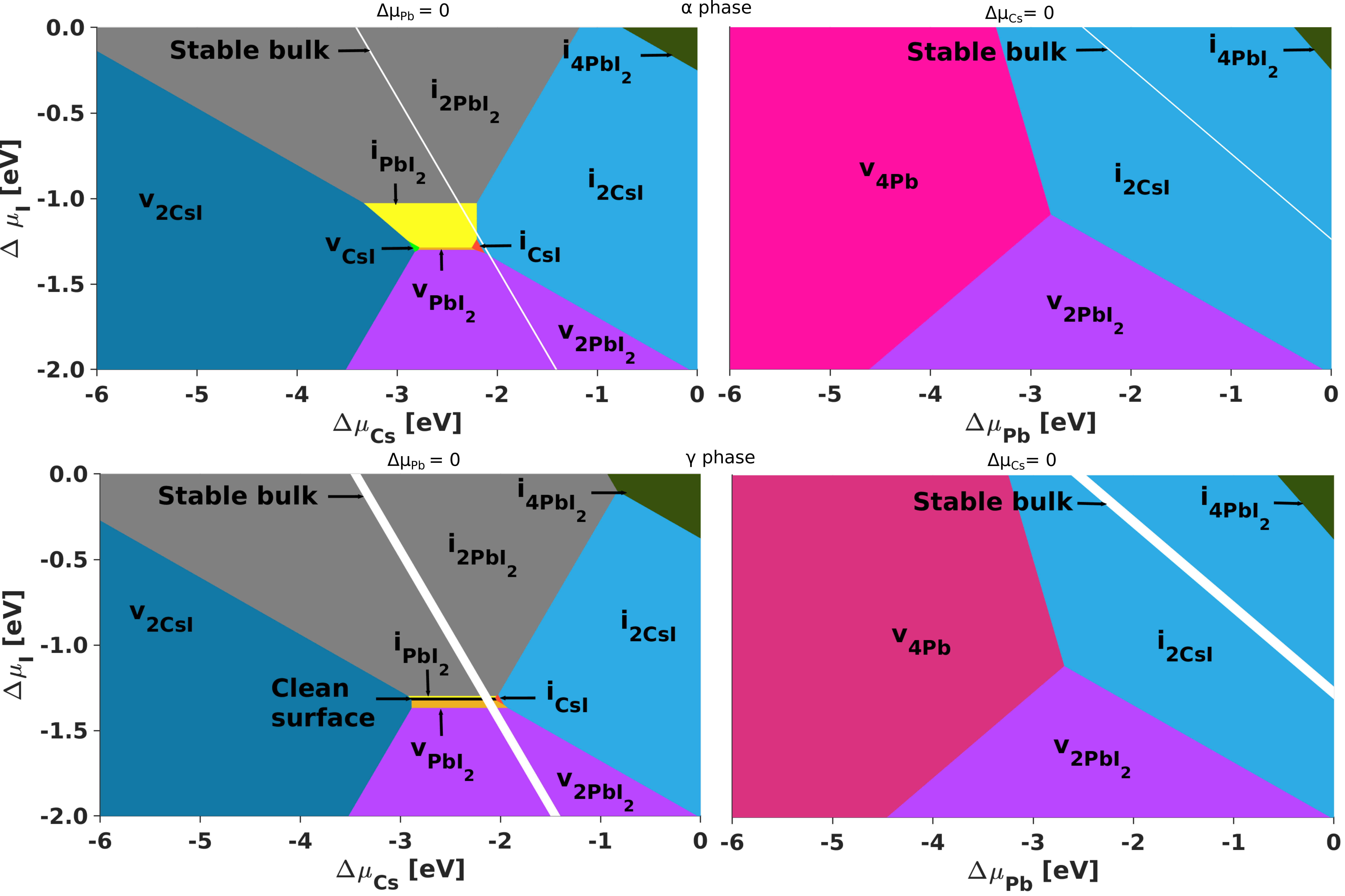}\vspace{-1.0em}
\caption{Surface phase diagrams of reconstructed $\text{PbI}_2^{}$-T models  with missing- and add-atoms. The white regions depict the thermodynamically stable range for the growth of bulk $\text{CsPbI}_3^{}$.}\label{spds_2}
\end{figure}

Figure~\ref{spds_2} shows the surface phase diagrams (SPDs) of PbI$_2$-T surfaces. On the left and right panels are SPDs of the $\upalpha$ and the $\upgamma$ phases at $\Delta\mu_{\text{Pb}}=0$ and $\Delta\mu_{\text{Cs}}=0$ respectively. Clearly, the clean surface is the stable. For instance, in the $\upalpha$ phase, we did not observe the clean surface (Fig.~\ref{spds_2} light top panel) and in the $\upgamma$ phase where the clean surface was observed, it only covers a thin layer of Fig.~\ref{spds_2} (left bottom panel). As discussed for CsI-T surfaces in Sec.~III~A~1, surface models that intersect the stable bulk are the most relevant. Here, v$_{2\text{PbI}_2^{}}$, i$_{\text{PbI}_2^{}}$, i$_{2\text{PbI}_2^{}}$  and i$_{4\text{PbI}_2^{}}$ are the most relevant models in $\upalpha$ phase. Similarly, the most relevant models in $\upgamma$ phase are v$_{2\text{PbI}_2^{}}$, v$_{2\text{PbI}_2^{}}$, i$_{\text{PbI}_2^{}}$, i$_{2\text{PbI}_2^{}}$  and i$_{4\text{PbI}_2^{}}$ as well as the clean surface. 

\newpage

\subsection{Relaxed geometries of \texorpdfstring{$\text{PbI}_2^{}$}{}-T models}

\begin{figure}[!htp]
\centering
\includegraphics[width=\linewidth]{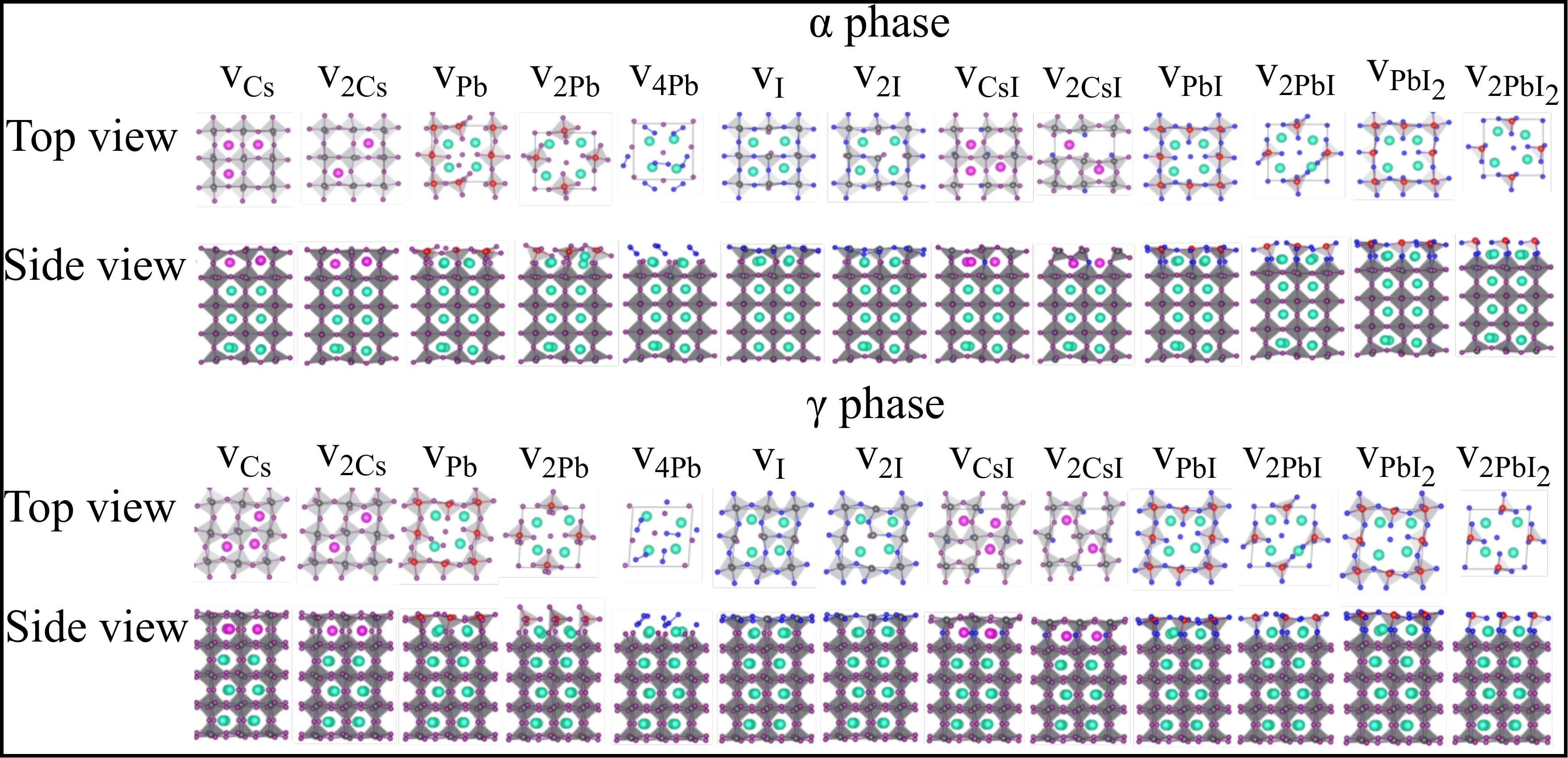}\\\vspace{0.5em}
\includegraphics[width=\linewidth]{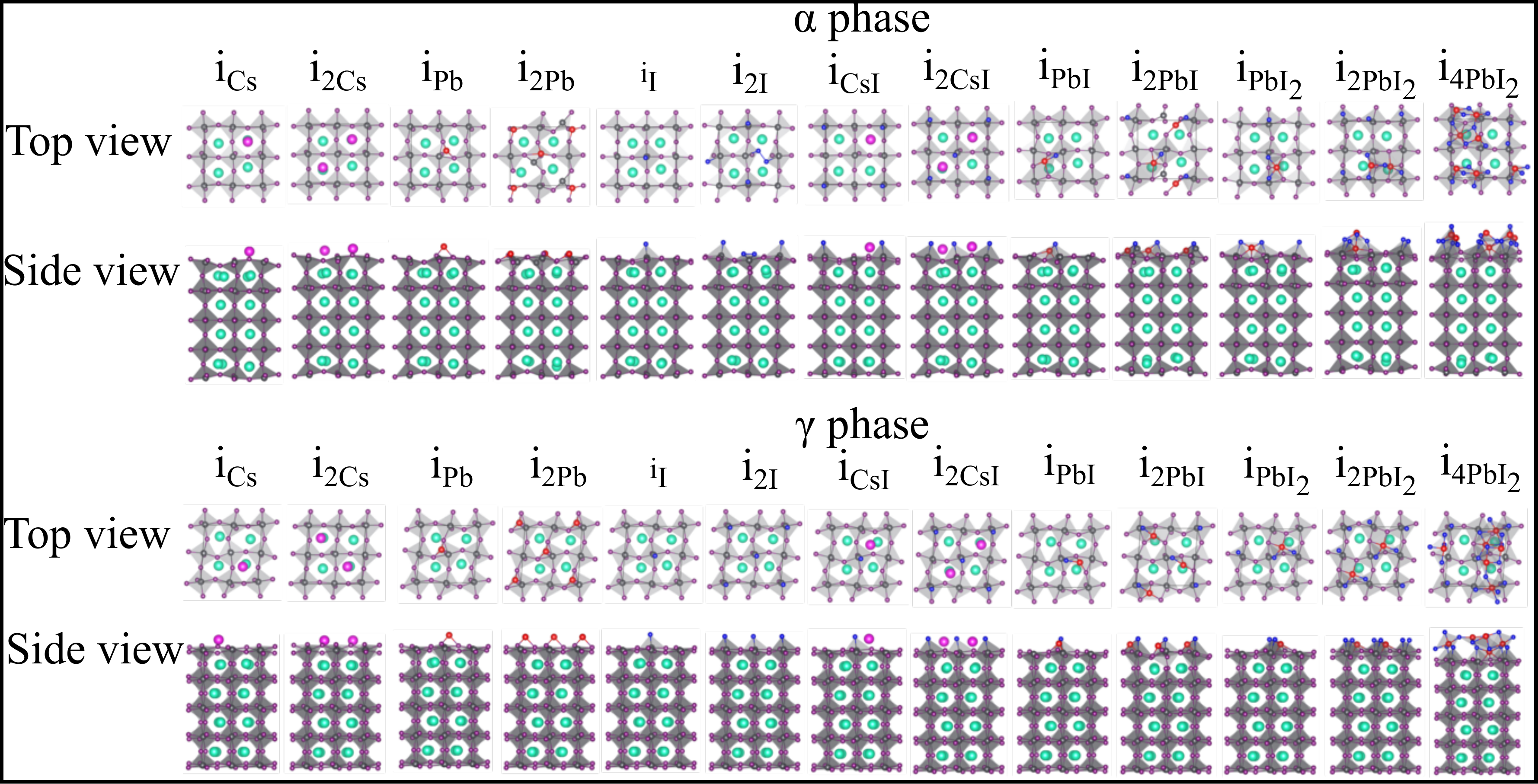}
\vspace{-1.0em}
\caption{Reconstructed surface structures with missing and add-atoms of $\text{PbI}_2^{}${}-T models in the $\upalpha$ and the $\upgamma$ phases.}\label{rel_pbi2}
\end{figure}

Reconstructed structures of PbI$_2$-T surface models are depicted in Fig.~\ref{rel_pbi2}. The top panel shows  models with missing atoms (complexes) while the bottom panel shows models with adatoms (complexes) in both phases. We observe slight shifts in atomic positions due to missing and add atoms. These shifts also resulted in octahedra tilting of the surface octahedra (Fig.~\ref{rel_pbi2}). We also observe I-migration in v$_{2\text{Pb}}$ for both phases. Here, two surface I-atoms move close to each other to form I$_2$ molecule with I-I bond length of $\sim2.8$ {\AA} in both $\upalpha$ and $\upgamma$ phases.

\newpage

\section{Formation energies of all studied surface-reconstruction models}

Table~\ref{energies} shows the formation energies at $\Delta\mu_{x} = 0$ ($x=\text{Cs, Pb and I}$) of all surface models in this work. Clearly, only surface models with the low formation energies were dominant in our surface phase diagrams. 

\begin{table}[!htp]
\caption{Formation energies (in eV) of CsI-T and $\text{PbI}_2^{}$-T surface models in the $\upalpha$ and the $\upgamma$ phases.} \label{energies}
\begin{tabular}{l@{\hspace{1em}}r@{\hspace{1em}}r@{\hspace{2em}}l@{\hspace{1em}}r@{\hspace{1em}}r} \hline\hline
\multicolumn{1}{c}{CsI-T} & \multicolumn{1}{c}{$\upalpha$} & \multicolumn{1}{c}{$\upgamma$} & \multicolumn{1}{c}{PbI$_2$-T} & \multicolumn{1}{c}{$\upalpha$} & \multicolumn{1}{c}{$\upgamma$} \\ \hline

v$_{\text{Cs}}$        &   $3.21$ &   $3.68$ & v$_{\text{Cs}}$        &   $3.59$ & $4.06$\\

v$_{2\text{Cs}}$       &   $7.27$ &    $7.7$ & v$_{2\text{Cs}}$       &   $7.86$ & $8.26$ \\

v$_{4\text{Cs}}$       &  $15.33$ &  $13.44$ & v$_{\text{Pb}}$        &   $1.57$ & $2.26$ \\

v$_{\text{Pb}}$        &   $2.64$ &   $3.05$ & v$_{2\text{Pb}}$       &   $4.57$ & $5.29$ \\

v$_{2\text{Pb}}$       &   $4.67$ &   $5.77$ & v$_{4\text{Pb}}$       &   $6.04$ & $6.28$ \\

v$_{\text{I}}$         &   $2.29$ &   $2.71$ & v$_{\text{I}}$         &   $2.01$ & $2.31$ \\

v$_{2\text{I}}$        &   $5.16$ &   $5.74$ & v$_{2\text{I}}$        &   $4.32$ & $4.82$ \\

v$_{\text{CsI}}$       &   $3.78$ &   $4.09$ & v$_{\text{CsI}}$       &   $3.72$ & $4.28$ \\

v$_{2\text{CsI}}$      &   $7.75$ &   $8.08$ & v$_{2\text{CsI}}$      &   $7.86$ & $8.41$ \\

v$_{\text{PbI}}$       &   $3.05$ &   $3.39$ & v$_{\text{PbI}}$       &   $2.30$ & $2.86$ \\

v$_{2\text{PbI}}$      &   $5.29$ &   $6.37$ & v$_{2\text{PbI}}$      &   $4.26$ & $4.73$ \\

v$_{\text{PbI}_2^{}}$  &   $3.47$ &   $3.63$ & v$_{\text{PbI}_2^{}}$  &   $2.23$ & $2.64$ \\

v$_{2\text{PbI}_2^{}}$ &   $6.95$ &   $7.49$ & v$_{2\text{PbI}_2^{}}$ &   $4.83$ & $5.37$\\ 

i$_{\text{Cs}}$        &  $-0.49$ &  $-0.36$ & i$_{\text{Cs}}$        &  $-1.81$ & $-1.25$ \\

i$_{2\text{Cs}}$       &  $-0.28$ &   $0.35$ & i$_{2\text{Cs}}$       &  $-2.72$ & $-2.23$ \\

i$_{\text{Pb}}$        &   $2.14$ &   $1.96$ & i$_{\text{Pb}}$        &   $0.99$ & $1.33$\\

i$_{2\text{Pb}}$       &   $2.74$ &   $3.13$ & i$_{2\text{Pb}}$       &   $1.66$ & $2.79$ \\

i$_{\text{I}}$         &  $-0.06$ &   $0.34$ & i$_{\text{I}}$         &  $-0.26$ & $0.25$ \\

i$_{2\text{I}}$        &  $-1.07$ &  $-0.68$ & i$_{2\text{I}}$        &  $-0.91$ & $0.60$\\

i$_{\text{CsI}}$       &  $-2.76$ &  $-2.69$ & i$_{\text{CsI}}$       &  $-3.88$ & $-3.37$ \\

i$_{2\text{CsI}}$      &  $-6.26$ &  $-5.97$ & i$_{2\text{CsI}}$      &  $-7.33$ & $-6.69$ \\

i$_{4\text{CsI}}$      & $-13.12$ & $-12.69$ & i$_{\text{PbI}}$       &  $-0.71$ & $-0.16$ \\

i$_{\text{PbI}}$       &   $0.17$ &   $0.47$ & i$_{2\text{PbI}}$      &  $-1.45$ & $-0.87$ \\

i$_{2\text{PbI}}$      &  $-1.04$ &   $0.97$ & i$_{\text{PbI}_2^{}}$  &  $-2.91$ & $-2.62$\\

i$_{\text{PbI}_2^{}}$  &  $-2.22$ &  $-1.95$ & i$_{2\text{PbI}_2^{}}$ &  $-4.97$ & $-5.22$ \\

i$_{2\text{PbI}_2^{}}$ &  $-4.21$ &  $-3.83$ & i$_{4\text{PbI}_2^{}}$ &  $-8.45$ & $-8.96$ \\ \hline\hline
\end{tabular}
\end{table}

%